# Current-driven dynamics and ratchet effect of skyrmion bubbles in a ferrimagnetic insulator

Saül Vélez[1,5,*], Sandra Ruiz-Gómez[2,3,6], Jakob Schaab[1], Elzbieta Gradauskaite[1], Martin S. Wörnle[4], Pol Welter[4], Benjamin J. Jacot[1], Christian L. Degen[4], Morgan Trassin[1], Manfred Fiebig[1] & Pietro Gambardella[1,*]

[1] Department of Materials, ETH Zurich, 8093 Zurich, Switzerland

[2] Departamento de Física de Materiales, Universidad Complutense de Madrid, 28040 Madrid, Spain

[3] Alba Synchrotron Light Facility, 08290, Cerdanyola del Valles, Barcelona, Spain

[4] Department of Physics, ETH Zurich, 8093 Zurich, Switzerland

[5] Present address: Condensed Matter Physics Center (IFIMAC), Instituto Nicolás Cabrera, and Departamento de Física de la Materia Condensada, Universidad Autónoma de Madrid, 28049 Madrid, Spain

[6] Present address: Max Planck Institute for Chemical Physics of Solids, 01187 Dresden, Germany

[*]e-mail: saul.velez@uam.es (S.V.); pietro.gambardella@mat.ethz.ch (P.G.)

**Magnetic skyrmions are compact chiral spin textures that exhibit a rich variety of topological phenomena and hold potential for developing high-density memory devices and novel computing schemes driven by spin currents. Here, we demonstrate room temperature interfacial stabilization and current-driven control of skyrmion bubbles in the ferrimagnetic insulator $Tm_3Fe_5O_{12}$ (TmIG) coupled to Pt. We track the current-induced motion of individual skyrmion bubbles. The ferrimagnetic order of the crystal together with the interplay of spin-orbit torques and pinning determine the skyrmion dynamics in TmIG and result in a strong skyrmion Hall effect characterized by a negative deflection angle and hopping motion. Further, we show that the velocity and depinning threshold of the skyrmion bubbles can be modified by exchange coupling TmIG to an in-plane magnetized $Y_3Fe_5O_{12}$ layer, which distorts the spin texture of the skyrmions and leads to a directional-dependent rectification of their dynamics. This effect, which is equivalent to a magnetic ratchet, is exploited to control the skyrmion flow in a racetrack-like device.**

Magnetic skyrmions are intensively investigated owing to their topological spin texture[1,2,3,4,5] and sensitive response to electric currents in thin-film devices[4,5,6,7]. These properties, which can be tuned by interface engineering, open unprecedented prospects for the development of skyrmion racetrack memories[4,8] and non-conventional logic devices[9,10]. Despite this surge of interest, skyrmion-based devices have so far only been realized in metallic systems.

Controlling the formation and dynamics of skyrmions in magnetic insulators (MIs) is a requirement for enabling low-power spintronic and magnonic applications[11,12,13]. Among different classes of MIs, rare-earth garnets coupled to heavy metals have opened exciting possibilities for electrically driving and detecting magnon currents over large distances[12,14,15] and for driving domain walls at high-speed[16,17] due to their low damping and ferrimagnetic order. Current-driven domain wall motion in centrosymmetric MIs is possible thanks to the interfacial Dzyaloshinskii-Moriya interaction (DMI)[16,17,18,19,20], which favours the formation of homochiral Néel walls[17]. These walls have the largest mobility in response to the spin-orbit torques (SOTs) generated by current flow in coupled heavy metal layers[21,22,23,24]. Recent measurements of the topological Hall effect in TmIG/Pt provided indirect evidence of the formation of skyrmion-like magnetic textures in thin-film MIs[20,25,26,27]. The possibility to vary the DMI, magnetic anisotropy, angular momentum, and damping of the rare-earth garnets by composition and strain engineering[19,20,25,28] makes these systems extremely appealing for tuning the skyrmion properties[29,30].

Despite this progress, the direct imaging and current-induced manipulation of skyrmions in MIs remain to be demonstrated. We thus lack crucial information on the nucleation, current-driven dynamics, and stability of skyrmions in MIs and how they compare to metallic systems. In this work, we demonstrate interfacial-stabilization and current-driven motion of skyrmion bubbles in a ferrimagnetic garnet at room temperature. We provide first insights into the skyrmion nucleation process, current-induced dynamics in the presence of pinning and thermal diffusion, skyrmion Hall angle, and rectification effects. Our results demonstrate the potential of MIs for hosting skyrmions and tuning their motion, and highlight the limitations that need to be addressed to realize efficient skyrmion devices.

**Sample design**

Our samples are $Y_3Fe_5O_{12}$(YIG)/$Tm_3Fe_5O_{12}$(TmIG)/Pt trilayers grown on (111)-oriented $Gd_3Sc_2Ga_3O_{12}$ (GSGG) with thickness of 10, 20, and 5 nm for YIG, TmIG, and Pt, respectively. Current lines were patterned in the shape of Hall bars by optical lithography and etching of Pt (see Fig. 1a), leaving the YIG/TmIG layers unetched (Methods). Magnetic characterization confirms that the TmIG and YIG layers exhibit out-of-plane and in-plane magnetic anisotropy, respectively (Supplementary

Note 2), with TmIG having a smaller coercivity and remanent magnetization (Fig. 1b) compared to films directly grown on GSGG[16,17,18,31]. This indicates a reduction of the magnetic anisotropy of TmIG due to the exchange coupling with YIG, which favours the formation of skyrmions without compromising their stability with temperature[26,29,30]. Unless otherwise specified, the YIG film was demagnetized before starting the measurements.

**Chirality and current-induced nucleation and motion of skyrmion bubbles**

Figure 1c shows magneto-optical Kerr effect (MOKE) microscopy images of the field-induced magnetization reversal of TmIG, which is characterized by the nucleation and expansion of labyrinthine stripe-like domains as commonly observed in garnets with out-of-plane magnetization[28]. By applying a current pulse, bubble domains nucleate out of a homogenous magnetic texture (Fig. 1d, upper image) as well as from the stripe domains (Fig. 1d, center and bottom images) in the region covered by the Pt current line. These bubble domains, which have an estimated radius $R$ of $0.5 - 1$ μm (Extended Data Fig. 1), are stable in time and are observed for magnetic fields $|H_z| \lesssim 25$ Oe, indicating that they are robust magnetic configurations. The best conditions to observe isolated bubbles correspond to a magnetic field $|H_z| = 20$ Oe, for which the magnetization is close to saturation and domains have not yet formed (point "1" in Fig. 1b). Subsequent application of current pulses results in the motion of both bubbles and stripe domains in the direction of the current, accompanied by the elongation or contraction of stripe domains and the nucleation of bubbles from stripe domains (Fig. 1e,f). This directional motion is the result of the current-induced SOTs that are exerted on the walls that delimit the domains[6,22,23].

According to the sign of the torques and spin Hall effect in TmIG/Pt[16,17,31], we conclude that the domain walls of both bubbles and stripes have a right-handed Néel chirality. The closed geometry of the bubbles and the chirality of the domain walls point to a skyrmionic texture[2,6], supporting previous reports of the topological Hall effect in TmIG/Pt[20,25,26,27]. Scanning nitrogen-vacancy (NV) magnetometry measurements of YIG/TmIG/Pt (Methods) confirm the skyrmionic nature of the bubbles. Figure 2 presents the stray field and reconstructed magnetization texture of a circular (Fig. 2a-c) and a deformed bubble due to pinning (Fig. 2d-f). The fits of the magnetic stray field reveal a domain wall width $\Delta_{DW} \sim 60$ nm and confirm the right-handed Néel chirality of the walls. Analysis of the stray field of straight domains in YIG/TmIG further suggests that the right-handed Néel chirality of the walls is also favored in Pt-free regions (Supplementary Note 3), indicating that the YIG/TmIG interface has a positive DMI, unlike GSGG/TmIG[17]. As the top TmIG/Pt interface has a positive DMI[17], we conclude that both YIG and Pt contribute to stabilize right-handed Néel walls and skyrmions in TmIG.

The threshold DMI strength required to stabilize chiral domain walls is given by[16,17,21] $D_{th} = 2\mu_0 M_s^2 t \ln 2 /\pi^2$, where $M_s$ and $t$ are the saturation magnetization and thickness of the magnetic layer, respectively. For our TmIG film (Supplementary Note 2), we obtain $D_{th} \sim 12$ µJ m$^{-2}$, a value that is compatible with the interfacial DMI found in TmIG-based heterostructures[16,17,18,19,26] ($D_{th}$ further reduces if demagnetizing fields are taken into account[32]). Interestingly, the low $M_s$ of TmIG lowers $D_{th}$ by two orders of magnitude compared to ferromagnets, evidencing the potential of ferrimagnetic garnets for stabilizing chiral structures even with weak DMI.

**Skyrmion Hall effect**

Measurements of current-driven displacements of isolated skyrmion bubbles show that they exhibit a skyrmion Hall effect[33,34] (Fig. 3), i.e., a transverse deflection relative to the current. The deflection direction depends on the magnetic polarity of the bubble's core (Fig. 3a), confirming that the skyrmion bubbles carry a topological charge[2]. Interestingly, the sign of the deflection angle $\phi_{sk}$ is opposite to that encountered in metallic ferromagnets. We ascribe the sign of $\phi_{sk}$ to the net positive angular momentum of our TmIG films, $s_{net} = -\frac{M_s}{\gamma} > 0$, where $\gamma$ is the effective gyromagnetic ratio (Methods and Supplementary Note 4). In metallic ferrimagnets, the reduced net angular momentum typically results in a small deflection angle[35,36], but this is not the case of TmIG, for which $|\phi_{sk}| \sim 40°$ (Fig. 3b). In the absence of disorder, the deflection angle is given by[35] (Supplementary Note 5)

$$\phi_{sk} \sim \tan^{-1}\left(-\frac{s_{net}}{s_{tot}}\frac{2\Delta_{DW}Q}{\alpha R}\right), \tag{1}$$

where $s_{tot}$ is the total angular momentum, $\alpha \sim 0.01$ the magnetic damping parameter[37], $\Delta_{DW} \sim 60$ nm and $R \sim 0.6$ µm (Fig. 2 and Extended Data Fig. 1), and $Q = +1/-1$ the topological charge for a skyrmion with core magnetization pointing up/down[2]. In our films we estimate $\frac{s_{net}}{s_{tot}} \sim -0.06$ (Methods), which leads to a large $|\phi_{sk}| \sim 50°$ due to the relatively low damping of TmIG. Another remarkable difference compared to metallic systems is that, in these, pinning strongly influences the deflection angle in the vicinity of the depinning threshold, resulting in a strong dependence of $\phi_{sk}$ on the current density[33,36,38]. This is not the case of TmIG as the average deflection angle is $|\bar{\phi}_{sk}| \sim 40°$ once the depinning threshold is reached (Fig. 3b). We emphasize that this result holds when averaging over several skyrmion trajectories. As discussed below, although disorder is lower than in polycrystalline metal films, the skyrmion motion in TmIG is strongly affected by pinning and thermal diffusion.

**Skyrmion trajectories and pinning effects**

We now determine whether the skyrmion trajectories are deterministic or stochastic and investigate pinning effects. Figure 4a reports the trajectories of 35 different non-interacting bubbles following the injection of a series of current pulses of density $J_x \sim 2 \times 10^{11}$ A m$^{-2}$ (Methods). Clearly, the displacements are neither linear in time nor in space. Further, histograms of the single-pulse displacements $\delta x$ and $\delta y$ along the $x$ and $y$ directions reveal a bimodal statistical distribution, which consists of a narrower peak centered at $\delta x, \delta y = 0$ and a broader one centered at $\delta x = x_0$ and $\delta y = y_0$ (Fig. 4b). The two modes of the distribution capture pulse events that did not lead to net bubble displacements (blue bars) and pulses that led to a net displacement (red bars), respectively. This behaviour indicates that the dynamics of the skyrmion bubbles driven by current pulses is strongly influenced by pinning at structural or magnetic defects. Because individual bubbles alternate between pinned and unpinned states, and because the probability of depinning is $P < 1$, the skyrmion bubbles move in the creep regime[38,39].

Upon depinning, the bubbles preferably move towards the direction set by the driving force and the Magnus force (Fig. 3 and Supplementary Fig. 8). The analysis of their shape further reveals that the bubbles tend to deform in the direction of motion as well as perpendicular to it, indicating that both SOTs and the skyrmion Hall effect concur in the deformation process in the presence of pinning[35,40,41,42] (Extended Data Fig. 2 and Supplementary Note 7). In most cases, however, the deformation is less than 10% relative to the circular shape, and only in less than 5% of the events clear deformations are observed (Fig. 2c,f and Extended Data Fig. 2a).

The individual longitudinal and transversal displacements have standard deviations $\sigma_{x_0}, \sigma_{y_0}$ larger than the mean $x_0, y_0$ values (Fig. 4b), indicating that the net skyrmion motion is accompanied by random hopping between pinning sites. We attribute this behaviour to the influence of disorder and current-induced thermal fluctuations on the displacements[9]. Measurements of the current threshold $J_x^{\mathrm{th}}$ for bubble depinning show that $J_x^{\mathrm{th}}$ decreases strongly upon increasing the pulse length $t_{\mathrm{p}}$, indicating that both SOTs and thermal effects concur in the depinning process (Extended Data Figs. 3 and 4). Applying stronger currents or longer pulses results in thermal motion dominating over directional motion and in the nucleation and annihilation of skyrmions[9,43,44], which prevent us from driving the system into the flow regime as reported for metallic systems[7,34,38,41,45].

The analysis of the mean displacements shows that the bubbles move, on average, by an amount that increases with $t_{\mathrm{p}}$ (Fig. 4c,d and Extended Data Fig. 5). However, the displacements tend to finite values $x_0 \approx 100$ nm and $y_0 \approx 70$ nm as $t_{\mathrm{p}}$ drops below $\sim 50$ ns (Fig. 4c,d), which is unexpected. These values are independent of the amplitude and direction of the current as well as of the YIG magnetization (Extended Data Fig. 6). Thus, they likely reflect a characteristic length scale of TmIG, namely the average hopping distance between pinning sites. This idea is supported by the distribution of bubble deformations, which is consistent with an average distance between pinning centers on the order of 100 nm (Fig. 2d-f and Supplementary Note 7). Because the bubble radius is larger than such a distance, the skyrmion dynamics is influenced by the pinning of the skyrmion wall to more than one pinning site. We also considered inertial and automotion effects[46,47] to explain the finite bubble displacements observed as $t_p \rightarrow 0$, but these effects appear unlikely in view of the properties of our system (Supplementary Note 8).

The hopping motion has strong consequences on the mean skyrmion velocity $\bar{v}_{sk} = \sqrt{(\overline{\Delta x})^2 + (\overline{\Delta y})^2}/t_p$ calculated, as customary in skyrmion studies, using the mean bubble displacements $\overline{\Delta x}$ and $\overline{\Delta y}$ averaged over all pulses, including those that did not lead to bubble motion. As shown in Fig. 4e, $\bar{v}_{sk}$ increases from about 2 m s$^{-1}$ at $t_p > 50$ ns to about 10 m s$^{-1}$ at $t_p = 10$ ns. This increase is attributed to the reduction of the thermal load on TmIG at shorter pulses, which reduces the random hopping and leads to a more efficient directional flow of the skyrmions. In these conditions, $\bar{v}_{sk}$ increases towards the flow regime limit, which we estimate as ~30 m s$^{-1}$ for $J_x = 1 \times 10^{11}$ A m$^{-2}$ (Methods). This limit, however, is hard to reach given that $J_x^{th}$ increases with decreasing $t_p$ (Extended Data Fig. 3).

It is known from experimental[33,34,35,38,40,43] and theoretical studies[38,39,42] that disorder strongly impacts the current-driven motion of skyrmions. Our findings reveal that the behaviour of skyrmions in a MI exhibits remarkable differences compared to metallic heterostructures. First, the density of pinning sites in TmIG, estimated from $(x_0 y_0)^{-1}$ for $t_p \rightarrow 0$ is $\approx 10^{-4}$ nm$^{-2}$, two orders of magnitude lower than in polycrystalline metal films[43]. Despite the lower disorder, the $J_x - t_p$ parameter space for skyrmion motion in TmIG is reduced to a narrow range due to current-induced heating dominating over the SOTs (Extended Data Figs. 3 and 4). We attribute this limitation to the exponential dependence of the skyrmion diffusion with temperature, which is expected to alter the skyrmion dynamics in materials with low damping and low disorder[9], preventing the use of currents strong enough to reach the flow regime. The thickness of our films also reduces the SOT driving force in comparison to ultrathin metallic systems[38,41,43].

Another difference with respect to metal films[33,34,38,41,43] is that the average skyrmion Hall angle $\bar{\phi}_{sk} = \tan^{-1}\left(\frac{\overline{\Delta x}}{\overline{\Delta y}}\right)$ in TmIG is very large (Figs. 3 and 4f), even though the skyrmions move in the creep regime. Theoretical models of skyrmions interacting with random point defects[39] or a granular magnetic anisotropy background[38,42,43] predict a decrease of $\bar{\phi}_{sk}$ with disorder, leading to $\bar{\phi}_{sk}$~0 in the creep regime. In TmIG, however, the density of defects is low compared to sputtered metal films, resulting in $\bar{\phi}_{sk}$ close to the flow limit given by Eq. 1, which only reduces from ~42° to ~34° upon decreasing $t_p$ from 100 to 10 ns (Fig. 4f). This reduction is another indication that the effects of pinning become more evident as $t_p$ is reduced. For long pulses, disorder has a small influence on the average skyrmion deflection, as inferred from the large $\bar{\phi}_{sk}$, but the thermal fluctuations are large, resulting in a reduction of $\bar{v}_{sk}$ (Fig. 4e). Conversely, for short pulses the thermal fluctuations reduce, leading to a more efficient SOT motion but also to a stronger influence of pinning on the skyrmions' trajectories, which results in the decrease of $\bar{\phi}_{sk}$ with decreasing $t_p$ (Fig. 4f). We also find that $\bar{\phi}_{sk}$ is nearly independent on $J_x$ (Fig. 3b), which we attribute to the competing action of SOTs and heating as $J_x$ increases above the depinning threshold. Future experimental and computational studies should aim at elucidating the influence of thermal diffusion[9] on the current-induced dynamics of skyrmions in materials with low damping and low disorder.

## Skyrmion ratchet effect

We now investigate the influence of the in-plane magnetization of YIG ($\mathbf{M}_{YIG}$) on the skyrmion dynamics in TmIG. The skyrmion trajectories remain affected by pinning in a homogenous $\mathbf{M}_{YIG}$, and the average deflection angle is about the same for different $\mathbf{M}_{YIG}$ configurations (Figs. 4a,b and 5). However, the depinning probability and the mean bubble displacements depend strongly on the orientation of $\mathbf{M}_{YIG}$ relative to $\mathbf{J}_x$ (Fig. 5b,d). In particular, the bubble displacements with $J_x > 0$ are much larger for $\mathbf{M}_{YIG}$ pointing along $-\mathbf{y}$ (Fig. 5a,b) than for $+\mathbf{y}$ (Fig. 5c,d), with the demagnetized case lying in between the two (Fig. 4a,b). This asymmetry is only observed when $\mathbf{M}_{YIG}$ is perpendicular to $\mathbf{J}_x$, with the motion of the bubbles being more (less) efficient when $\mathbf{J}_x \times \mathbf{M}_{YIG} \sim -\mathbf{z}(+\mathbf{z})$ regardless of their topological charge (Supplementary Table 1). These observations suggest that $\mathbf{M}_{YIG}$ modifies the escape probability of the bubbles from the pinning potential, whereas the density of pinning sites is not significantly influenced by $\mathbf{M}_{YIG}$. We attribute this escape asymmetry to the distortion of the magnetic texture of the bubbles' wall induced by the exchange coupling with YIG, which results in SOTs of different strength depending on the orientation of $\mathbf{M}_{YIG}$ relative to $\mathbf{J}_x$. The mechanism that we propose can be explained as follows. In the absence of current, the magnetic moments in the wall of a bubble ($\mathbf{m}_{DW}$) tilt towards $\mathbf{M}_{YIG}$ (Fig. 6a,b). In the presence of current, $\mathbf{m}_{DW}$ acquires an additional tilt in the direction of the damping-like SOT ($\mathbf{T}^{DL}$)[24,48], such that $d\mathbf{m}_{DW}/dt \propto -\mathbf{T}^{DL} \propto J_x \mathbf{y}$ (Fig. 6c,d). Therefore, the distortion produced by $\mathbf{J}_x$ opposes (favours) the one induced by $\mathbf{M}_{YIG}$ when $\mathbf{J}_x \times \mathbf{M}_{YIG}$ points towards $-\mathbf{z}(+\mathbf{z})$, an asymmetry that is consistent with the experiments (Fig. 5) and holds for both signs of $Q$ (Supplementary Note 10). Thus, given that $\mathbf{T}^{DL}$ controls both the depinning and the displacement of the bubbles, and that $\mathbf{T}^{DL} \propto \mathbf{m}_{DW} \times (\mathbf{m}_{DW} \times \mathbf{y})$ is proportional to the $x$-component of $\mathbf{m}_{DW}$, the skyrmion bubbles move more (less) efficiently when the magnetic tilt towards $\mathbf{y}$ is minimal (maximal). Importantly, no asymmetry in the skyrmion dynamics is observed when $\mathbf{M}_{YIG}$ and $\mathbf{J}_x$ are collinear (Extended Data Fig. 7), in agreement with our model. In addition, no changes of $\phi_{sk}$ are expected due to a change in $\mathbf{T}^{DL}$, which is also consistent with our observations (Figs. 4a and 5a,c).

We next exploit the asymmetry in the current-driven skyrmion depinning and displacements induced by $\mathbf{M}_{YIG}$ to rectify the skyrmion motion. In the vicinity of the depinning threshold, the asymmetry of $\mathbf{T}^{DL}$ induced by $\mathbf{M}_{YIG}$ leads to the unidirectional motion of the skyrmions. Figures 6e,f and 6g,h show the trajectory of a few skyrmion bubbles for alternating sequences of positive and negative current pulses with $\mathbf{M}_{YIG}$ pointing along $-\mathbf{y}$ and $+\mathbf{y}$, respectively. Clearly, the bubbles move only for one polarity of the current, which depends on the orientation of $\mathbf{M}_{YIG}$. This skyrmion ratchet effect is similar to a magnetic gate, which can be used to induce a net skyrmion displacement from random current excitations or for preventing the skyrmions to move along a particular direction when using an alternating current to generate SOTs. An analogous ratchet effect is observed for stripe domains due to the homochiral nature of the domain walls in TmIG (Supplementary Fig. 11). We remark that, whereas the ratchet effect requires the presence of pinning, the directional asymmetry of the skyrmion motion due to $\mathbf{M}_{YIG}$ does not.

**Conclusions**

We showed that skyrmion bubbles can be stabilized and driven by proximity charge currents in a centrosymmetric MI coupled to Pt. Despite the reduced density of defects in TmIG, we find that pinning and thermal skyrmion diffusion severely affect the motion of the skyrmion bubbles, which are constrained in the creep regime for currents below the emergence of thermal instabilities. The bubbles move by intermittent sequential jumps between nearby pinning sites, which result in a broad distribution of longitudinal and transverse displacements. Remarkably, the skyrmion Hall effect is large and opposite compared to ferromagnets, which we ascribe to the relatively low damping, low density of defects, and positive angular momentum of the TmIG film. In principle, the ferrimagnetic order of MIs allows for controlling the sign and amplitude of the net angular momentum, and therefore tune the skyrmion Hall effect[35] and mobility[49]. Future realizations of skyrmions in MIs should aim at improving their thermal stability for a broader range of currents, possibly by a concomitant increase of the magnetic anisotropy and DMI in rare earth garnets[19,20,25,26]. Moreover, pinning effects should be minimized to achieve deterministic and efficient skyrmion motion. Finally, we demonstrated control over the skyrmion dynamics by exchange coupling TmIG to an in-plane magnetized YIG layer. As the skyrmion's driving force depends on the polarity of the current relative to the magnetization of the in-plane layer, it is possible to rectify the skyrmion's motion along a predefined direction. This new aspect of the skyrmion dynamics provides an additional tool for tailoring the mobility of skyrmions in spintronic devices.

## Acknowledgements

We acknowledge André Thiaville, Aleš Hrabec, and Christoforos Moutafis for useful discussions and Marvin Müller for technical assistance with the MOKE setup. This work was funded by the Swiss National Science Foundation (Grants No. 200020-200465 P.G., 200021-188414 M.T., 200021-178825 M.F., PZ00P2-179944 B.J.J., and 200020-175600 C.L.D.), by the European Research Council (Advanced Grant 694955-INSEETO M.F.), and by ETH Zürich (Career Seed Grant SEED-20 19-2 S.V.). S.R. acknowledges support from the Spanish Ministry of Economy and Competitiveness (FPI fellowship and Grant No. RTI2018-095303-B-C53). S.V. acknowledges financial support by the Ministry of Science, Innovation and Universities through the 'Maria de Maeztu Program for Units of Excellence in R&D (Grant No. CEX2018-000805-M) and by the Comunidad de Madrid through the Atracción de Talento program (Grant No. 2020-T1/IND-20041).

## Author contributions

S.V. conceived the study and coordinated the experimental work. J.S., E.G., and M.T. grew and characterized the films. S.V. fabricated the devices. S.V. and S.R. performed the transport and MOKE experiments and analysed the data. B.J.J. assisted with time-resolved transport measurements. M.S.W. and P.W. performed the NV measurements with the help of S.V. S.V. and P.G. wrote the manuscript. P.G., M.T., C.L.D., and M.F. supervised the work. All authors contributed to the scientific discussion and manuscript revisions.

## Competing interests

The authors declare that they have no competing interests.

**FIGURES**

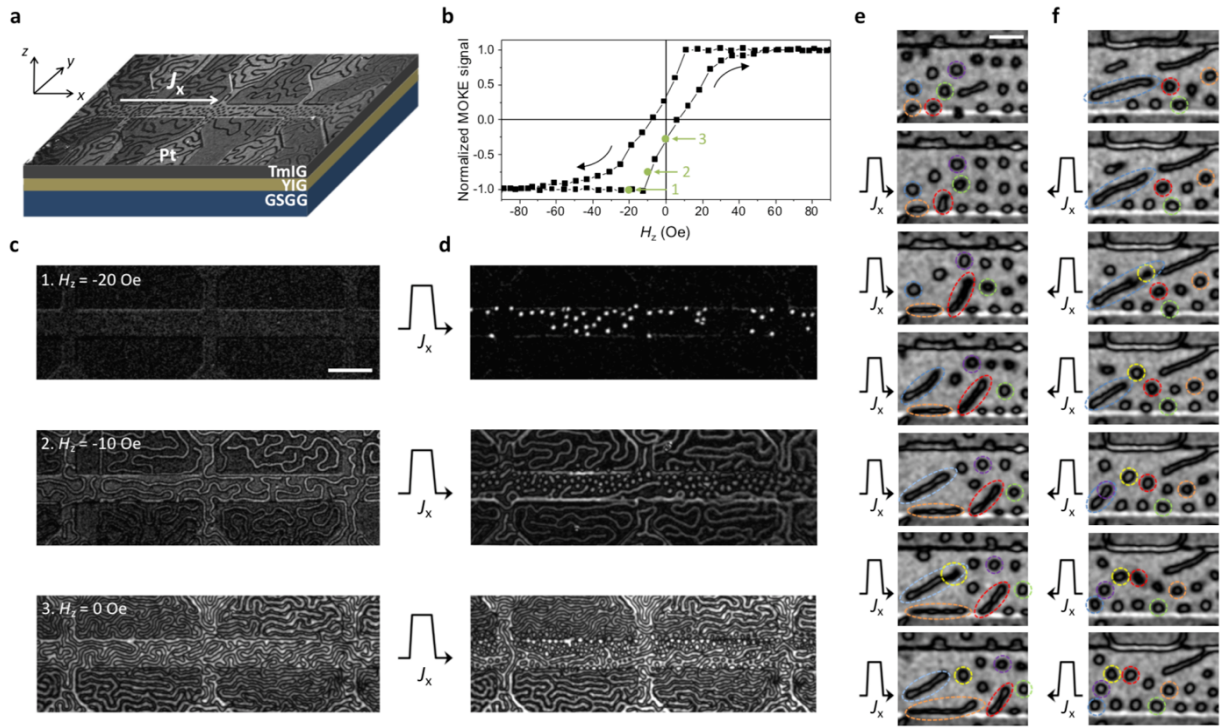

**Figure 1 | Current-driven nucleation and dynamics of bubble domains in TmIG. a**, Schematic of the device structure with a superimposed wide-field differential MOKE image showing magnetic domains in TmIG. Bright (dark) contrast corresponds to regions with up (down) magnetization. The coordinate axis and the current line are indicated. **b**, Magnetic hysteresis loop of TmIG measured by MOKE microscopy in the region covered by the current line while sweeping the out-of-plane magnetic field $H_z$. The films were first demagnetized by cycling the in-plane field in loops of alternating polarity and decreasing amplitude. **c**, Sequence of differential MOKE images taken at different magnetic fields (green dots in **b**) showing that the magnetization reversal proceeds by the formation of labyrinthine stripe-like domains. **d**, MOKE images taken after the application of a current pulse $J_x = 6.2 \times 10^{11}$ A m$^{-2}$ with length $t_p = 40$ ns to the domain structures shown in **c**, revealing the nucleation of bubble domains (top image) as well as the nucleation and breaking of the stripe domains into bubble domains (center and bottom images). All magnetic configurations in **c** and **d** remained stable after one hour. **e**, **f**, From up to down, sequence of MOKE images showing snapshots of the current-driven dynamics for a sequence of positive and negative current pulses (see arrow direction), respectively. The domains move along the direction of the current and their dynamics is influenced by pinning, resulting in the deformation of bubbles into stripes or vice versa and the formation of new bubbles from stripes. Dashed lines of different colours evidence the position of selected domains before and after pulsing. We attribute domain motion outside the current line to repulsive dipolar interactions. The measurements are performed with $H_z = -10$ Oe, $|J_x| = 3 \times 10^{11}$ A m$^{-2}$, and $t_p = 40$ ns. Scale bars in **c**, **e** are 15 and 5 µm, respectively.

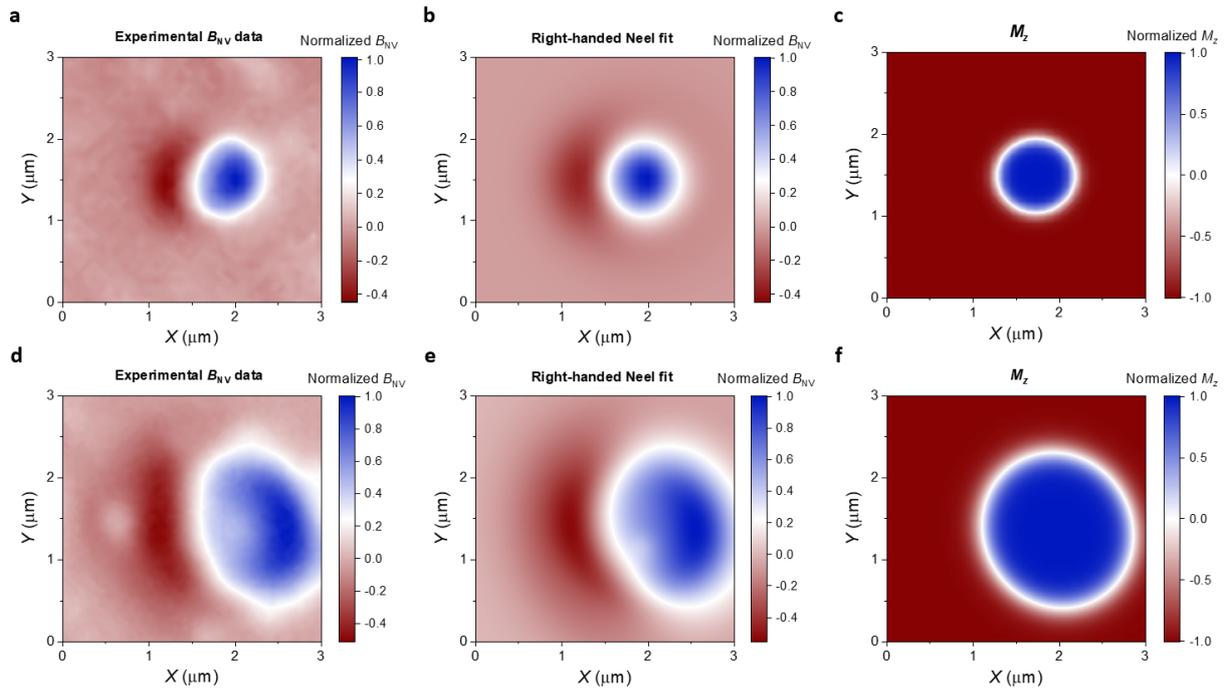

**Figure 2 | Nitrogen-vacancy magnetometry analysis of the skyrmion bubbles in TmIG. a**, Normalized stray field map $B_{\mathrm{NV}}(X, Y)$ of a single skyrmion bubble in YIG/TmIG/Pt measured by scanning the NV tip over the $XY$ plane. **b**, Best fit of the data in **a**, which corresponds to a circular bubble with diameter $2R \sim 950$ nm and a right-handed Néel wall with $\Delta_{\mathrm{DW}} = 60$ nm. **c**, Reconstructed out of plane magnetic component $M_z$ corresponding to the data in **b**. **d-f**, Same as in **a-c** for a deformed skyrmion bubble due to wall pinning. Fit of the data in **d** assuming an ellipsoidal shape of the bubble with arbitrary orientation in the $XY$ plane. The best fit is obtained for a right-handed Néel wall with $\Delta_{\mathrm{DW}} = 60$ nm and radial axes 940 and 830 nm. $H_z = -20$ Oe in **a** and **d**. See Supplementary Note 3 for details regarding the fitting procedure.

**a**

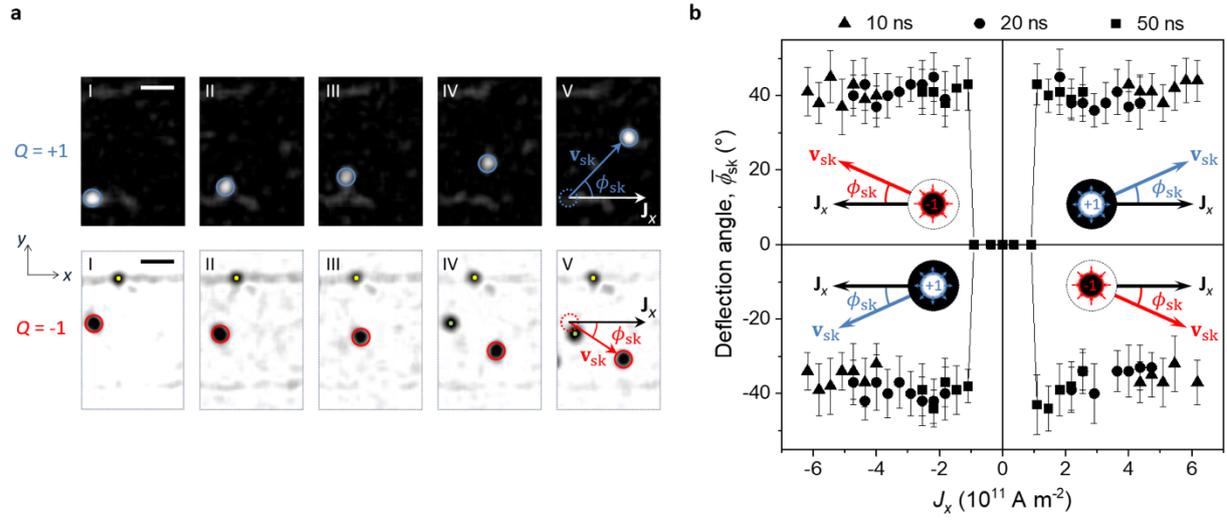

**Figure 3 | Skyrmion Hall effect in a ferrimagnetic insulator. a**, Sequence of differential MOKE images showing the position of isolated skyrmion bubbles during a series of positive current pulses. The top and bottom images correspond to bubbles with $Q = +1$ and $Q = -1$, respectively. The direction of $\mathbf{J}_x$, $\mathbf{v}_{sk}$ and $\phi_{sk}$ are indicated in the rightmost image of each sequence. The images are selected from a sequence of current pulses with $J_x = 2.0 \times 10^{11}$ A m$^{-2}$ and $t_p = 20$ ns. Bright (dark) contrast corresponds to regions with up (down) magnetization. In the bottom images, the yellow spots indicate a skyrmion bubble trapped at a defect site and the green ones a bubble driven into the imaging region during the pulse sequence. The scale bars are 3 μm. **b**, Average skyrmion deflection angle $\bar{\phi}_{sk}$ measured for different amplitude, length, and direction of the current pulses for skyrmion bubbles with $Q = +1$ and $-1$. Each data point is an average performed over ten independent skyrmion bubbles with error bars representing the standard deviation (Methods). Different symbols correspond to different pulse lengths. Data points with $\phi_{sk} = 0$ indicate no single skyrmion depinning events. The lowest $|J_x|$ values shown for $t_p = 20$ and 10 ns indicate the current threshold for skyrmion depinning at these pulse lengths. The insets indicate the orientation of $\mathbf{J}_x$ and $\mathbf{v}_{sk}$, sign of $\phi_{sk}$, and the magnetic configuration of the skyrmion bubbles. $H_z = -20$ (+20) Oe for $Q = +1$ ($-1$).

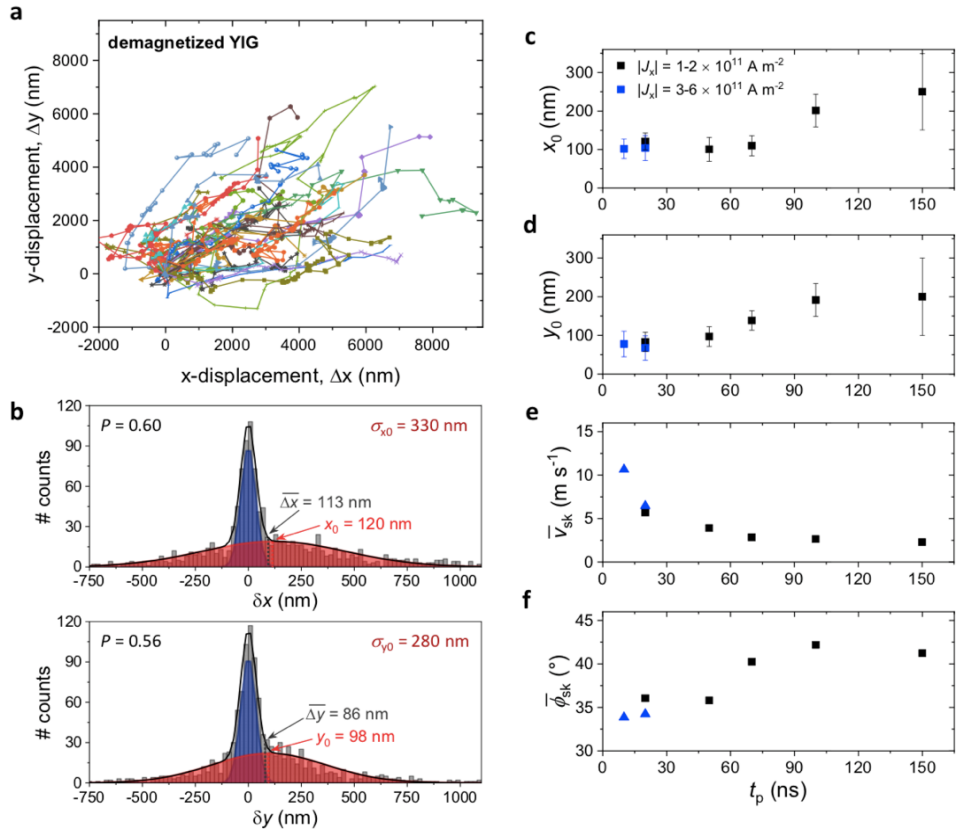

**Figure 4 | Statistical analysis of the trajectories of the skyrmion bubbles and pinning effects**. **a**, Representative data showing the trajectory of 35 different non-interacting skyrmion bubbles following the application of a sequence of current pulses with $J_x > 0$ for demagnetized YIG. The pulse length is $t_p = 50$ ns, the current density $J_x \sim 2 \times 10^{11}$ A m$^{-2}$, $H_z = -20$ Oe, and $Q = +1$. The initial position of all the trajectories is set at (0,0); each data point represents the position of a bubble after a current pulse, labelled by a different symbol and colour. **b**, Histogram of individual bubble displacements $\delta x$ and $\delta y$ computed as the difference in position $(x, y)$ between adjacent data points along the trajectories shown in **a**, representing bubble displacements per pulse. The grey bars represent the experimental data and the black lines the fits obtained using a double Gaussian distribution. The blue-shaded peak centered at $\delta x, \delta y = 0$ accounts for pulses that did not lead to bubble displacements. The red-shaded peak captures the distribution of pulses that led to bubble motion, centered at $\delta x = x_0$ and $\delta y = y_0$ with standard deviation $\sigma_{x_0}$ and $\sigma_{y_0}$. $P$ indicates the relative weight of the area under the red curve relative to the total histogram area; $\overline{\Delta x}$ and $\overline{\Delta y}$ are the average bubble displacements per pulse along $\boldsymbol{x}$ and $\boldsymbol{y}$, respectively, including all pulse events. **c-f**, Analysis of the skyrmion trajectories as function of $t_p$. **c**, **d**, Mean displacements $x_0$ and $y_0$ as a function of the pulse length; the error bars are the standard errors calculated from the variance of those parameters to the double Gaussian fit function. **e**, **f**, Pulse length dependence of the mean bubble velocity $\bar{v}_{sk}$ and Hall angle $\bar{\phi}_{sk}$ calculated from $\overline{\Delta x}$ and $\overline{\Delta y}$ (Extended Data Fig. 5). Different symbols in **c-f** indicate the current density.

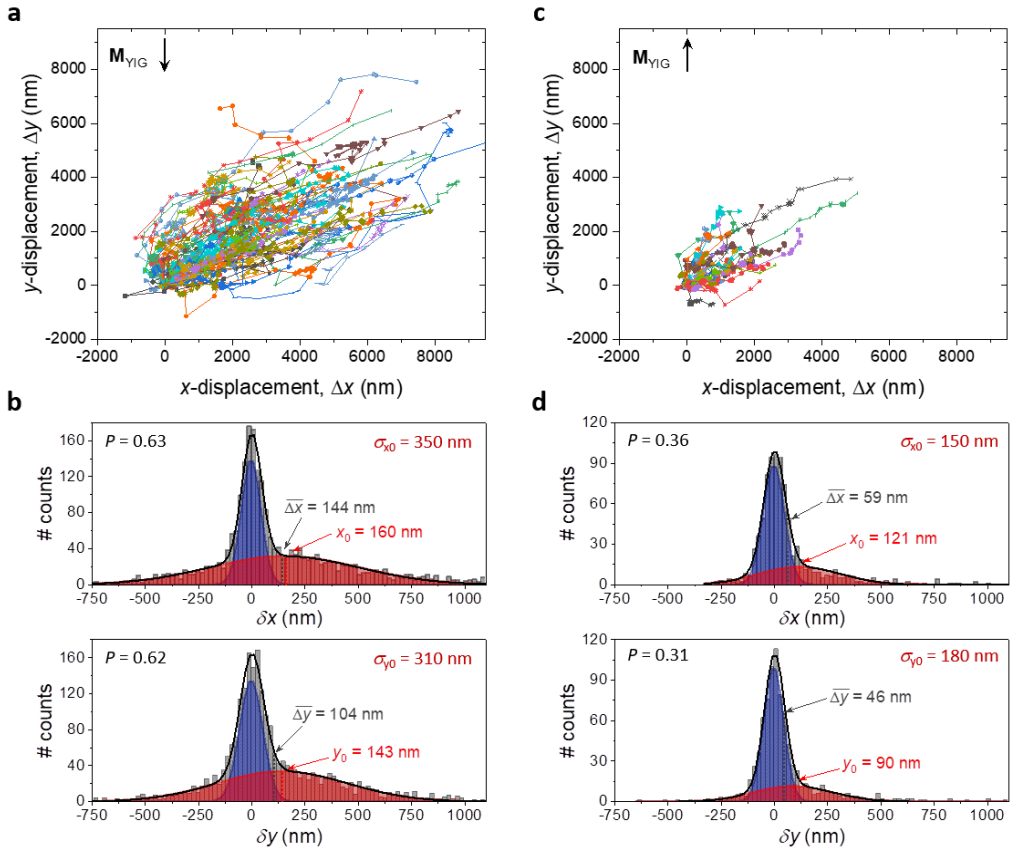

**Figure 5 | Statistical analysis of the trajectories of the skyrmion bubbles for opposite orientations of** $\mathbf{M}_{\text{YIG}}$. **a**, Representative data showing the trajectory of 68 different bubbles following the application of a sequence of current pulses $J_x > 0$ for $\mathbf{M}_{\text{YIG}} \parallel -\mathbf{y}$. **b**, Histogram of individual bubble displacements along the $x$ and $y$ directions extracted from the bubble trajectories shown in **a**. **c**, **d**, Same as **a**, **b** for 26 bubbles and $\mathbf{M}_{\text{YIG}} \parallel \mathbf{y}$. The pulse length is 50 ns, the current density $\sim 2 \times 10^{11}$ A m$^{-2}$, $H_z = -20$ Oe, and $Q = +1$. The orientation of $\mathbf{M}_{\text{YIG}}$ is set by a constant in-plane magnetic field $|H_y| = 10$ Oe. The grey bars in **b** and **d** represent the experimental data and the black lines the fits obtained using a double Gaussian distribution, separately represented in blue and red colours (see caption of Fig. 4a,b).

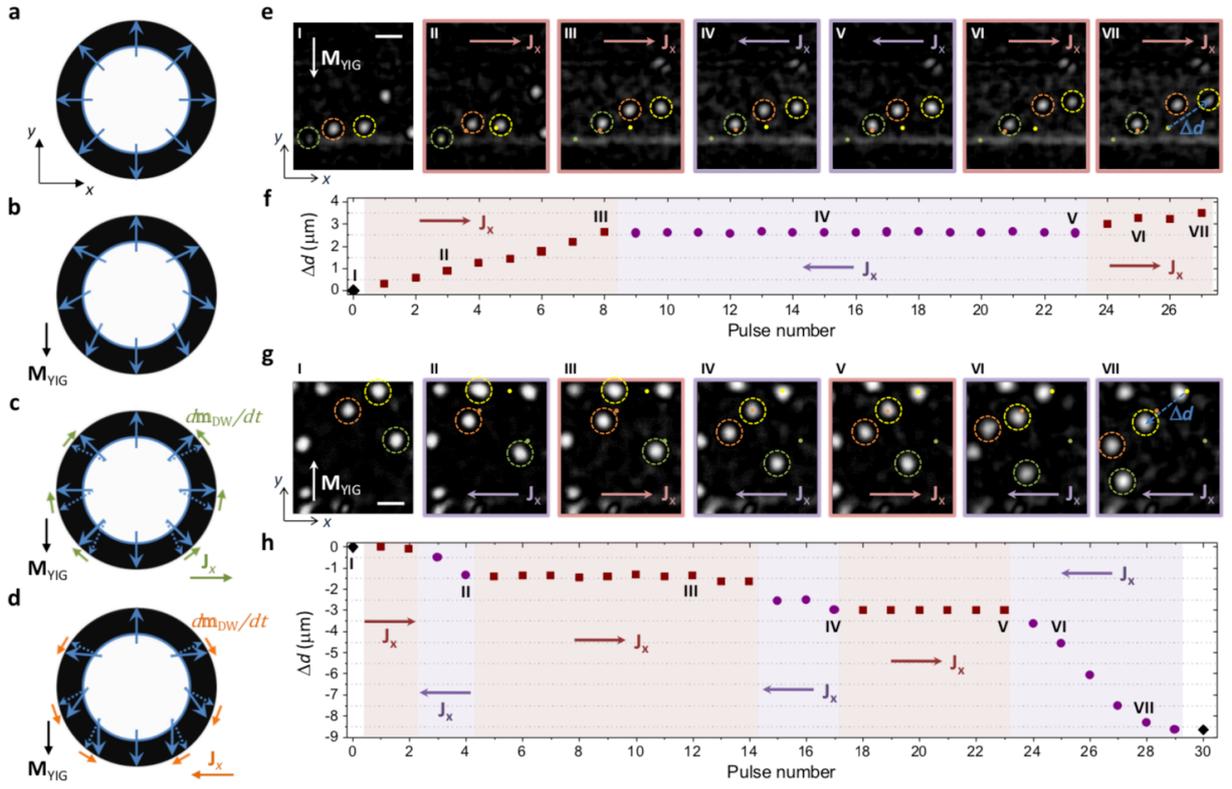

**Figure 6 | Skyrmion ratchet effect. a**, Schematic of the domain wall magnetic moments $\mathbf{m}_{DW}$ in a right-handed $Q = +1$ bubble. **b**, Distorted skyrmion bubble due to the exchange coupling with $\mathbf{M}_{YIG} = -M_{YIG}\boldsymbol{y}$. **c**, **d**, Additional distortion of the skyrmion bubble shown in **b** due to the dynamic action of the SOTs in TmIG/Pt. The dashed and solid arrows indicate $\mathbf{m}_{DW}$ before and during the pulse, respectively, with green and orange colours indicating the direction of $d\mathbf{m}_{DW}/dt$ for opposite current directions. The reduced (enhanced) distortion of $\mathbf{m}_{DW}$ in **c** (**d**) results in a larger (smaller) current-induced driving force. **e**, Representative images showing the displacement of $Q = +1$ skyrmion bubbles following the application of a sequence of current pulses of alternating polarity with $\mathbf{M}_{YIG}$ pointing towards $-\boldsymbol{y}$. The bubbles are labelled by dashed coloured circles; dots indicate the initial positions of the corresponding bubbles (see image #I). **f**, Total displacement $\Delta d = \sqrt{\Delta x^2 + \Delta y^2}$ averaged for the trajectory of the three skyrmion bubbles shown in **e** (see image #VII). The sign of $\Delta d$ corresponds to the sign of $\Delta x$. The direction of $\mathbf{J}_x$ and the corresponding image number are indicated in the pulse sequence. **g**, **h** Same as **e**, **f** but for $\mathbf{M}_{YIG}$ pointing towards $+\boldsymbol{y}$. $|J_x| = 1.8 \times 10^{11}$ A m$^{-2}$, $t_p = 50$ ns, and $H_z = -20$ Oe in **e**, **g**. The orientation of $\mathbf{M}_{YIG}$ is set by a constant in-plane magnetic field $|H_y| = 10$ Oe. Scale bars in **e**, **g** are 3 and 5 μm, respectively.

## Methods:

**Data availability.** The data that support the findings of this study have been deposited in the Research Collection database of the ETH Zurich and are available from https://doi.org/10.3929/ethz-b-000542503.

**Films growth and characterization.** The YIG and TmIG films were epitaxially grown onto (111)-oriented $Gd_3Sc_2Ga_3O_{12}$ (GSGG) substrates (lattice constant $a$ = 12.56 Å) by pulsed laser deposition to achieve high tensile strain (~2%) of the TmIG layer, which promotes TmIG to exhibit perpendicular magnetic anisotropy as earlier demonstrated[17,50]. The properties of the heterostructure were tuned by varying the thickness of the layers and the deposition conditions. For the sample investigated in this work, the YIG film was deposited at 720 °C with an oxygen background pressure of 0.15 mbar. The laser fluence was set to 0.9 J cm$^{-2}$, and the repetition rate at 2 Hz. The TmIG film was consecutively grown at 650 °C with an oxygen pressure of 0.2 mbar, and the laser fluence and repetition rate were set to 1.35 J cm$^{-2}$ and 8 Hz, respectively. After deposition, the sample was annealed at 750 °C under 120 mbar oxygen pressure for 30 min and cooled down to room temperature at a rate of -10 K/min in 200 mbar oxygen pressure. To ensure a high quality of the TmIG/Pt interface, the YIG/TmIG films grown were directly transferred to the sputter chamber without breaking vacuum, where the Pt layer was deposited at room temperature for 3 minutes at a power of 10 W in 0.05 mbar Ar plasma. The thickness of the layers was calibrated by X-ray reflectometry and determined to be 9.8, 20.2, and 5.1 nm for YIG, TmIG, and Pt, respectively. The uncertainty for the garnet layers is 2.0 nm. Atomic force microscopy measurements of the surface topography showed a root-mean-square roughness of about 0.18 nm over a 5 x 5 µm$^2$ area. The films were magnetically characterized in a superconducting quantum interference vibration sample magnetometer (SQUID-VSM) system. We estimate the saturation magnetization of TmIG and YIG to be 60 and 175 kA m$^{-1}$, respectively. See Supplementary Notes 1 and 2 for further details regarding the structural, topographic, and magnetic characterization of the films.

**Device fabrication.** The Pt layer was patterned into Hall bars (consisting of three Hall crosses separated by $L = 50$ µm and width $W = 10$ µm) by photolithography and subsequent Argon plasma etching following the recipe of Ref. 17, leaving the YIG/TmIG films unetched. The required etching time for removing the Pt layer was calibrated in reference films. 48 devices were patterned in each sample. Measurements were performed in 6 different samples, from which 2 showed skyrmion bubble stabilization and motion driven by electric currents. All data presented in this work was acquired in the same sample. The current-induced skyrmion motion measurements were performed in the same device with the exception of the data presented in Extended Data Figs. 6 and 7 and Supplementary Fig. 9. The NV data was acquired in a third device. Additional measurements were performed in other devices from the same sample, all showing same characteristics. The topography of the etched structures was characterized by atomic force microscopy, from which we determined the thickness of TmIG to be reduced by ~0.5 nm due to etching and the root-mean-square roughness in the etched region to be ~0.2 nm (Supplementary Fig. 2).

**MOKE measurements.** We used a home-built wide-field polar MOKE microscope with Koehler illumination to measure the out-of-plane component of TmIG[17]. As described in Ref. 17, the light source consists in a collimated light emitting diode from Prizmatix Ltd., model MIC-LED-455L, whose spectral emission is characterized by a maximum peak emission at 454 nm, centroid at 455 nm, and a full width at half maximum of 28 nm. The setup was equipped with two sets of orthogonal coils for the generation of out-of-plane and in-plane magnetic fields. For the skyrmion bubble generation and current-driven motion studies, current pulses were injected using an AGILENT 8114A (100V/2A) and a Kentech RTV40 sub-ns pulse generator. The impedance matching with the pulse generators was achieved by connecting a 50 $\Omega$ resistance in parallel to the Pt current line.

Magnetic contrast was enhanced by taking differential MOKE images, i.e., each image was subtracted by a reference image captured in a fully magnetized state as employed in Ref. 17. The contrast is defined as the absolute value of the differential image, being black (bright) when the difference is zero (maximal). A change of the magnetic field applied reduces the contrast due to the Faraday effect. Accordingly, the reference image for a fully down (up) magnetized state was taken at $H_z = -20$ (+20) Oe. Consequently, optimized contrast is observed in the upper images of Fig. 1c,d and the images of Figs. 3 and 6. Conversely, contrast is reduced for the center ($H_z = -10$ Oe) and bottom ($H_z = 0$ Oe) images of Fig. 1c,d as well as the images of Fig. 1e,f ($H_z = -10$ Oe). For the bottom images of Fig. 3a ($H_z = +20$ Oe; $Q = -1$), the contrast was inverted to preserve the colour-code definition black (bright) corresponding to down (up) magnetization.

For the current-driven skyrmion dynamics experiments (Figs. 3-6), the skyrmion bubbles were prepared as follows. First, we demagnetized the films by cycling the in-plane field in loops of alternating polarity and decreasing amplitude (a similar result can be obtained by cycling $\mathbf{H}_z$ instead). Second, for experiments with $\mathbf{M}_{YIG}$ polarized (Figs. 5,6), we then applied a $\mathbf{H}_y$ field. Third, we fully saturated the TmIG film up (down) by applying a magnetic field $H_z = +100$ (−100) Oe, and then decreased the magnetic field to $H_z = +20$ (−20) Oe. Fourth and last, we nucleated skyrmion bubbles by applying current pulses $J_x$ of amplitude and length between 5 and $8 \times 10^{11}$ A m$^{-2}$ and 20 to 40 ns, respectively, resulting in an apparent random nucleation of bubbles. The amplitude, length, polarity, and number of the pulses was adjusted to obtain a low-concentration of skyrmion bubbles (i.e., less than 1 bubble per $5 \times 5$ μm$^2$ device area). Bubble nucleation, however, is favoured at defect sites[40,44] as well as influenced by the current-induced Oersted field. Whereas the latter favours bubble nucleation at the side where the out-of-plane component of the Oersted field opposes $H_z$, the presence of defects can be inferred from the preferred nucleation of skyrmion bubbles at certain positions of the device. Moreover, we observe that the skyrmions located at a few of those spots remain trapped upon the application of current pulses. Therefore, we excluded those skyrmion bubbles for the analysis of the skyrmion dynamics.

For the acquisition of the skyrmion trajectories, differential MOKE images were taken at a rate of $\sim 1$ frame/pulse while the pulses were continuously applied at a repetition rate of 2 Hz. The trajectories of the bubbles were automatically determined by using a tracking software that binarizes the images and determines the position of the bubbles' center during the image sequence. The standard deviation of the blue distributions in Figs. 4b and 5b,d, which is about 50 nm, indicates that the position of the bubbles' center can be determined with sub-100 nm resolution. However, we remark that the internal structure and shape of the bubble's walls cannot be resolved at this scale due to the optical resolution of the system, which we estimate to be $\sim 300$ nm from measurements performed in a reference sample. For the evaluation of the skyrmion trajectories (Figs. 3-5), only mobile bubbles separated by at least 5 μm from neighbour bubbles were considered; in other words, we exclude from the analysis those bubbles that are permanently pinned at defect sites or that might interact with each other, as well as bubbles close to the edge of the track. For the statistical analysis, different skyrmion bubbles from different repeats and moving across different areas of the device were considered. Each data point in Fig. 3b corresponds to the average of over 10 independent bubbles with the deflection angle computed from the initial and final positions of the bubbles measured after a sequence of current pulses. All data points presented in Fig. 4c-f are extracted from the trajectory of a minimum of 50 and up to 250 skyrmion bubbles. Data taken for both polarities of $\mathbf{J}_x$ were considered for the analysis presented in Fig. 4c-f.

For the experiments of Figs. 5 and 6e-h, in addition to $H_z$, a permanent magnetic field $H_y = \pm 10$ Oe is applied to keep $\mathbf{M}_{\mathrm{YIG}}$ saturated along. For fields up to $|H_y| \sim 25$ Oe, condition at which the bubble domains transform into stripe domains, the impact of $\mathbf{H}_y$ on the skyrmion dynamics is found to be negligible compared to $\mathbf{M}_{\mathrm{YIG}}$, indicating that the ratchet effect arise from the exchange coupling with YIG. See Supplementary Note 9 for more details.

**Scanning NV magnetometry.** The stray field maps of domain walls and skyrmions in YIG/TmIG/Pt were acquired with a custom-built nanoscale scanning diamond magnetometer (NSDM) microscope[17]. This technique is based on a single NV defect located at the apex of a diamond tip. By scanning the tip over the $XY$ surface of the sample, one can sense the magnetic stray field $B_{\mathrm{NV}}(X, Y)$ emanating from the surface with high magnetic sensitivity and nanometer spatial resolution, from which the spin texture of domain walls and skyrmions can be determined[17,51,52,53]. Experiments were carried out in ambient environment by employing a commercial monolithic diamond probe tip from QZabre Ltd. (www.qzabre.com). The NV center spin resonance was monitored by optically detected magnetic resonance (ODMR) spectroscopy[54] using a nearby microwave antenna (~2.9 GHz) for spin excitation and fluorescence microscopy (520 nm excitation, 630–900 nm detection) for spin state readout. The orientation of the NV center and its stand-off distance with respect to the sample surface were characterized beforehand using a reference sample, with the latter determined to be $114 \pm 17$ nm. Due to the relatively large stray field of the film, measurements were conducted at an additional distance of 100 nm from the surface. Although this prevented us to resolve local features below 100 nm resolution, $\Delta_{DW}$ can be reliably fitted because the domain wall profile extends over a distance that is several times larger than the domain wall width (Supplementary Eq. 1). Measurements in Fig. 2 were performed on isolated bubble domains nucleated via the application of current pulses in the presence of a magnetic field $H_z = -20$ Oe. The straight domain in Supplementary Fig. 6 was nucleated by decreasing the field to $H_z = -15$ Oe from a fully down magnetized state. The field remained applied during the whole data acquisition. The details of the data analysis and determination of the magnetic texture of the domain walls and skyrmions are presented in Supplementary Note 3.

**Skyrmion Hall angle and velocity evaluation.** For the estimates of $\phi_{sk}$ and $v_{sk}$, we use the Wangsness relation[55] and set the magnetic moment and the Landé $g$ factors of the tetrahedral/octahedral $Fe^{3+}$ and the dodecahedral $Tm^{3+}$ sublattices of TmIG to be $M_{s,1} = 175$ kA m$^{-1}$ and $M_{s,2} = 115$ kA m$^{-1}$, and $g_1 = 2.0$ and $g_2 = 7/6$, respectively (see Refs. 56,57 and Supplementary Note 4). Accordingly, in our TmIG film $\frac{s_{net}}{s_{tot}} = \left(\frac{M_{s,1}}{g_1} - \frac{M_{s,2}}{g_2}\right) / \left(\frac{M_{s,1}}{g_1} + \frac{M_{s,2}}{g_2}\right) \approx -0.06$. Based on previous reports[17], and considering that the magnetic anisotropy of our film is lower, we estimate the domain wall width of TmIG to be $\Delta_{DW} \sim 50$ nm. This value is consistent with the wall width inferred from NV magnetometry measurements, which is estimated to be about 60 nm (Fig. 2 and Supplementary Note 3).

The skyrmion velocity in the flow regime is given by[38,45] $v_{sk} = -\frac{1}{\sqrt{1+\eta^2}} \xi_{DL} J_x \gamma \frac{\pi R}{4}$, where $\eta = -\frac{s_{tot}}{s_{net}} \frac{\alpha R}{2Q\Delta_{DW}} \sim 0.8$, $\xi_{DL}$ is the effective field per unit current density associated to the damping-like SOT, and $\gamma = g \frac{\mu_B}{\hbar}$, with $\mu_B$ the Bohr magneton, $\hbar$ the reduced Planck constant, and $g = \frac{M_s}{\left(\frac{M_{s,1}}{g_1} - \frac{M_{s,2}}{g_2}\right)} \approx -5.4$ in our TmIG film. According to previous reports of SOT efficiency in TmIG/Pt heterostructures[16,31,58], and considering the thickness of our films, we estimate $\xi_{DL} \sim 2 \times 10^{-15}$ T A$^{-1}$ m$^2$ in our devices. Accordingly, we estimate $v_{sk} \sim 35$ m/s for $|J_x| = 1 \times 10^{11}$ A m$^{-2}$. See Supplementary Note 5 for more details.

## Additional Information

**Supplementary Information** is available for this paper at https://doi.or/XXXX (to be inserted by the publisher).

# Extended Data Figures

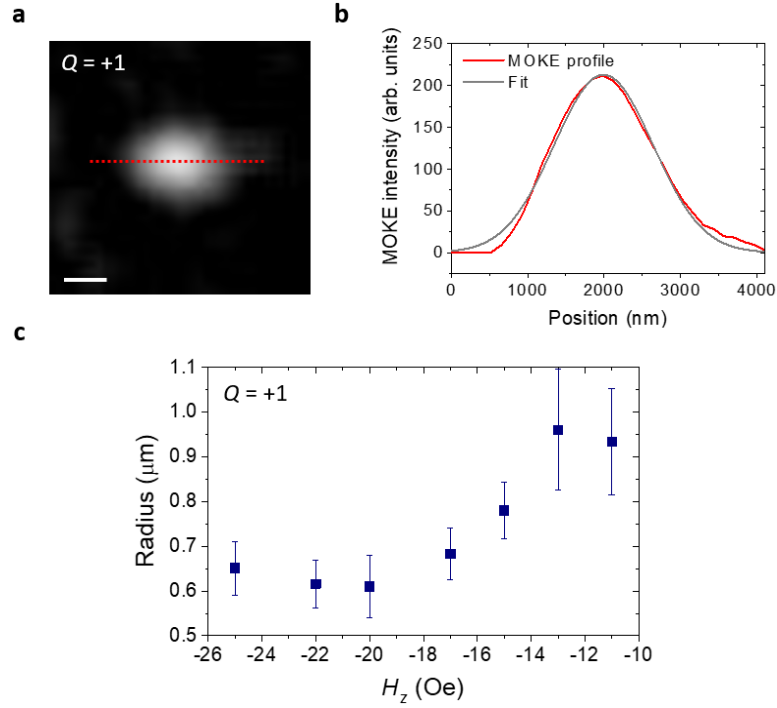

**Extended Data Figure 1 | Skyrmion bubble radius. a**, Differential MOKE image of a representative skyrmion bubble for $H_z = -20$ Oe. The white (dark) contrast indicates regions with **m** of TmIG pointing up (down). Scale bar, 1 μm. **b**, Line profile of the MOKE intensity taken along the red dashed line in **a** (red solid line) together with its fitting (light grey) assuming that the skyrmion is a square box function having an ellipsoidal shape convoluted by a Gaussian function with standard deviation ~300 nm, which represents the spatial resolution of the MOKE set up[49]. The domain wall width, which is $\Delta_{DW} \sim 60$ nm (Fig. 2), is neglected in the fitting procedure. The main diameters $a$ and $b$ of the skyrmion bubble are extracted by fitting the two orthogonal axes of the ellipsoid. From the fit in **b** we estimate $a \sim 1.2$ μm. **c**, Skyrmion bubble radius $R$ as a function of $H_z$. The radius is estimated as $R = (a + b)/4$. Each data point corresponds to the average value obtained from fitting over 10 independent skyrmion bubbles with circular shape ($b/a \gtrsim 0.9$). The error bars represent the standard deviation of the measurements. See Supplementary Note 6 for discussion on the field dependence of the skyrmion bubble radius and bubble stabilization with $H_z$.

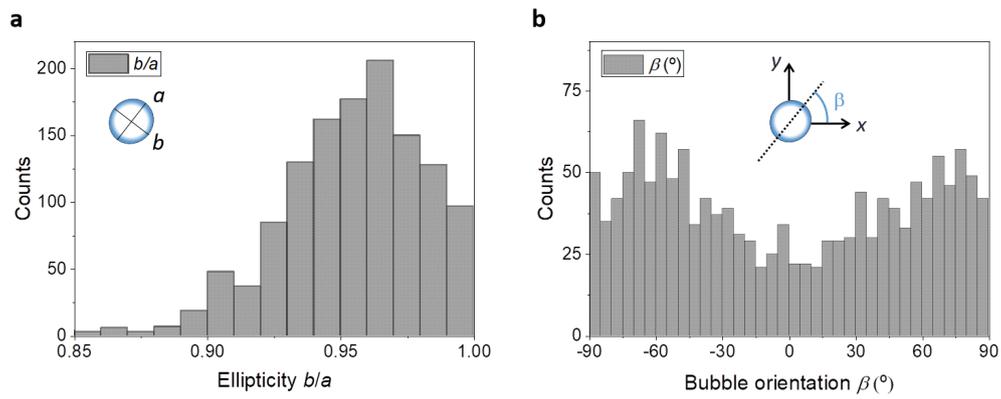

**Extended Data Figure 2 | Skyrmion ellipticity and orientation**. **a**, Statistical analysis of the skyrmion ellipticity $b/a$ from fitting over 1000 skyrmion bubbles assuming an ellipsoidal shape (see schematic). The histograms are extracted from analyzing the MOKE images of the bubble trajectories presented in Fig. 4a. **b**, Analysis of the orientation of the skyrmion ellipsoids from the data in **a**. The angle $\beta$ defines the orientation of the longest axis of the ellipsoid with respect to the current (see schematic).

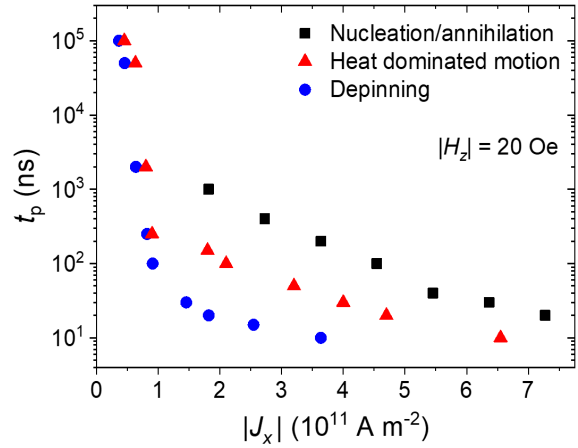

**Extended Data Figure 3** | $J_x - t_p$ **threshold conditions** for skyrmion depinning (blue circles), heat dominated motion (red triangles), and random nucleation and annihilation of skyrmion bubbles in TmIG (black squares). $|H_z| = 20$ Oe and the YIG layer is demagnetized. No difference was observed between $Q = +1$ and $-1$ skyrmion bubbles. Supplementary Note 9 presents the current threshold for skyrmion depinning in the presence of $\mathbf{H}_y$, i.e., with $\mathbf{M}_{YIG}$ controlled with an in-plane magnetic field. All measurements of the skyrmion dynamics presented in this work were performed for $t_p, |J_x|$ conditions comprised between the curves defined by the blue circles and the red triangles. Above the threshold defined by the red triangles, the mean displacements $\overline{\Delta x}, \overline{\Delta y}$ abruptly drop, indicating that the skyrmion dynamics are dominated by Joule heating induced random skyrmion motion rather than by SOTs. We attribute this behavior to the exponential increase of the skyrmion diffusivity with temperature, which is expected in materials with low disorder and low damping such as TmIG[9].

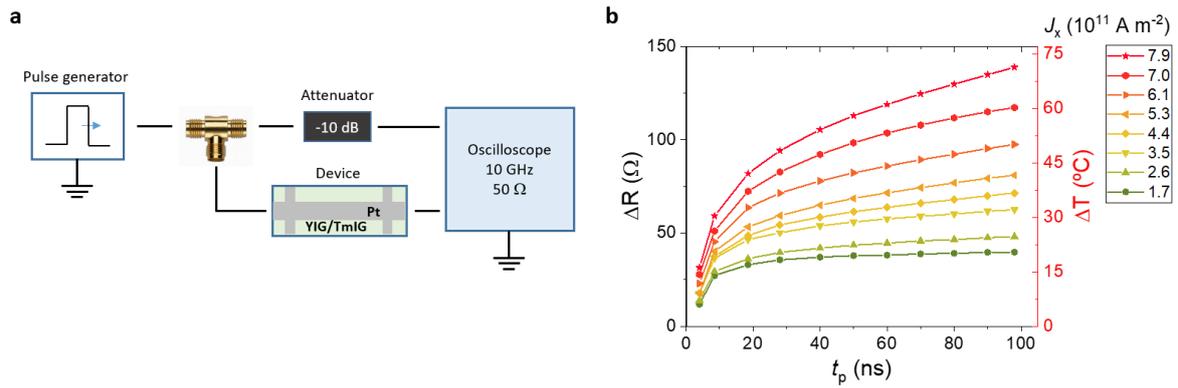

**Extended Data Figure 4 | Current-induced temperature increase. a**, Schematic of the experimental setup. The voltage output of the pulse generator is applied through the device and both the voltage drop at the device and the pulse are monitored with an Oscilloscope with an internal impedance of 50 Ω. From these measurements, we can precisely determine the evolution of the sample resistance during the current pulse. **b**, Increase of the sample resistance during the application of a current pulse $t_{\mathrm{p}} = 100$ ns (left axis). The pulse starts at $t = 0$. The colour indicates different set currents computed from the base resistance of the device $R_0(295\ \mathrm{K}) = 1130\ \Omega$. The right axis shows the increase of temperature calculated from calibration measurements. We estimate the threshold for heat dominated motion for temperature increases of about 20 K.

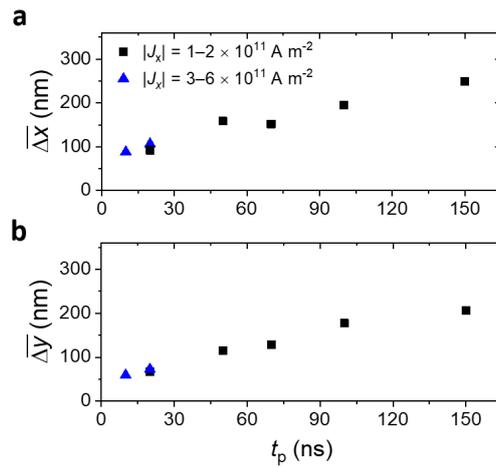

**Extended Data Figure 5 | Mean $\overline{\Delta x}$ and $\overline{\Delta y}$ displacements for YIG demagnetized. a, b,** Average bubble displacements per pulse $\overline{\Delta x}$ and $\overline{\Delta y}$ as a function of $t_p$ computed considering all pulse events, i.e., including those that did not lead to bubble displacements. $\overline{\Delta x}$ and $\overline{\Delta y}$ increase linearly with the pulse length, exhibiting a finite value as $t_p \to 0$. The difference between $\overline{\Delta x}$, $\overline{\Delta y}$ and $x_0, y_0$ (Fig. 4c,d) arises from the decrease of the bubble depinning probability when decreasing $t_p$. Different symbols indicate different current densities. $H_z = -20$ Oe ($Q = +1$). YIG is demagnetized.

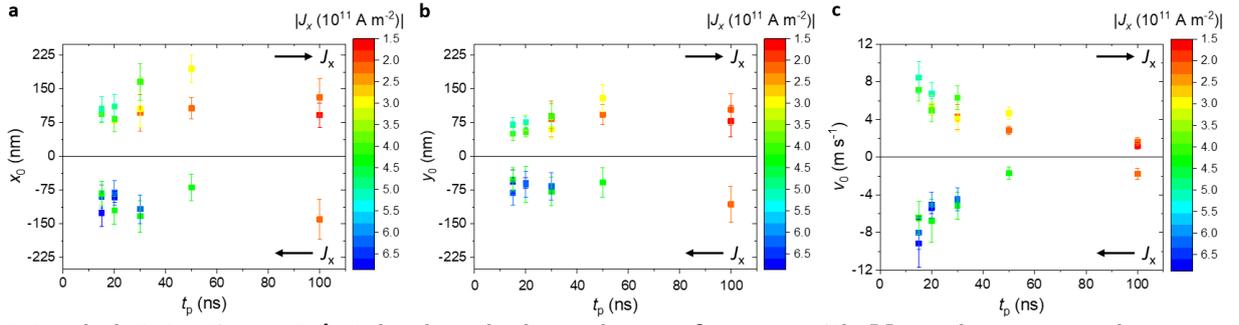

**Extended Data Figure 6 | Pulse length dependence of $x_0, y_0$ with $\mathbf{M}_{\mathrm{YIG}}$ along $-\mathbf{y}$. a, b,** Mean displacement values $x_0, y_0$ extracted from the trajectory of several skyrmion bubbles with $\mathbf{M}_{\mathrm{YIG}}$ along $-\mathbf{y}$ (see Fig. 4a,b for details regarding the analysis). Data taken for $H_z = -20$ Oe ($Q = +1$) and for both polarities of $\mathbf{J}_x$ (indicated by an arrow). The sign of $x_0, y_0$ corresponds to the sign of $\mathbf{J}_x$. Different colours indicate the current density. The error bars are the standard errors of $x_0, y_0$ calculated from the variance of these magnitudes to the double Gaussian distribution of $\delta x, \delta y$. As for the case of YIG demagnetized (Fig. 4c,d), $x_0$ and $y_0$ tend to finite values when $t_p \to 0$. Remarkably, the $x_0, y_0(t_p \to 0)$ values are similar for both directions of $\mathbf{J}_x$ and similar to the ones measured for YIG demagnetized. As $t_p$ increases, $x_0$ and $y_0$ start to increase from a pulse length threshold value that depends on the amplitude of the current. Larger (smaller) current densities are required for driving skyrmion bubbles with $\mathbf{M}_{\mathrm{YIG}}$ pointing to $-\mathbf{y}$ and $J_x < 0$ ($J_x > 0$), which is in agreement with the ratchet effect reported in Figs. 5 and 6 and Supplementary Note 10. **c,** Velocity of the skyrmion bubbles computed as $|v_0| = \sqrt{x_0^2 + y_0^2}/t_p$ from the data shown in **a** and **b**. The sign of the velocity is defined by the sign of $x_0$; the error bars are computed by error propagation. As observed for the mean bubble velocity $\bar{v}_{\mathrm{sk}}$ for YIG demagnetized (Fig. 4e), $v_0$ increase as $t_p$ reduces (note that $v_0(t_p)$ exhibits a stepper increase than $\bar{v}_{\mathrm{sk}}(t_p)$ when reducing $t_p$).

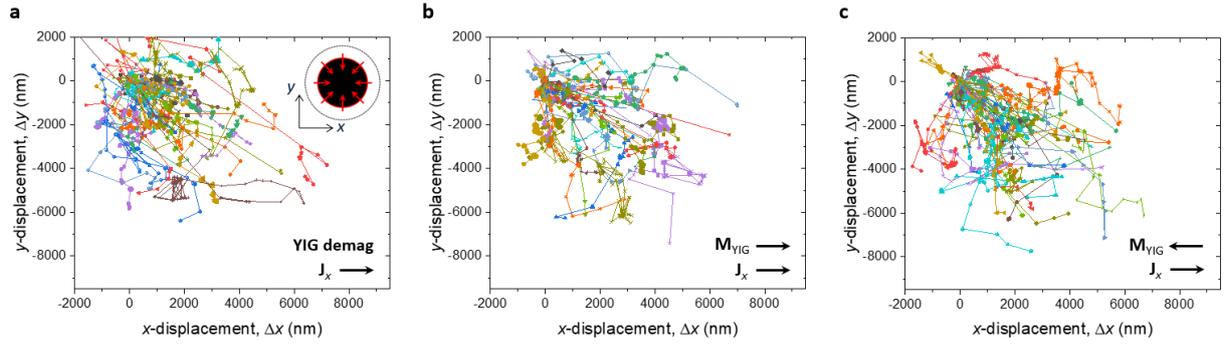

**Extended Data Figure 7 | Skyrmion trajectories with $\mathbf{M_{YIG}}$ collinear to $\mathbf{J_x}$. a**, Skyrmion trajectories with $Q = -1$ and YIG demagnetized (see inset's schematic). Data taken in a different device from the same YIG/TmIG/Pt heterostructure. $H_z = +20$ Oe, $J_x = 3.5 \times 10^{11}$ A m$^{-2}$, and $t_p = 20$ ns. The measurement protocol is the same employed in Fig. 4a. **b, c**, Skyrmion trajectories for $\mathbf{M_{YIG}}$ pointing to $+\boldsymbol{x}$ and $-\boldsymbol{x}$, respectively. $|H_x| = 10$ Oe. In contrast to the difference observed between $\mathbf{M_{YIG}}$ parallel to $+\boldsymbol{y}$ and $-\boldsymbol{y}$ (Figs. 5a and 5b), the skyrmion dynamics for $\mathbf{M_{YIG}}$ collinear with the current is, within the error, independent on the direction of $\mathbf{M_{YIG}}$. No clear differences between $\mathbf{M_{YIG}} \parallel \pm\boldsymbol{x}$ (**b,c**) and the demagnetized case (**a**) can be identified either.

# Supplementary Information for

# Current-driven dynamics and ratchet effect of skyrmion bubbles in a ferrimagnetic insulator

Saül Vélez[1,5,*], Sandra Ruiz-Gómez[2,3,6], Jakob Schaab[1], Elzbieta Gradauskaite[1], Martin S. Wörnle[4], Pol Welter[4], Benjamin J. Jacot[1], Christian L. Degen[4], Morgan Trassin[1], Manfred Fiebig[1] & Pietro Gambardella[1,*]

[1] Department of Materials, ETH Zurich, 8093 Zurich, Switzerland

[2] Departamento de Física de Materiales, Universidad Complutense de Madrid, 28040 Madrid, Spain

[3] Alba Synchrotron Light Facility, 08290, Cerdanyola del Valles, Barcelona, Spain

[4] Department of Physics, ETH Zurich, 8093 Zurich, Switzerland

[5] Present address: Condensed Matter Physics Center (IFIMAC), Instituto Nicolás Cabrera, and Departamento de Física de la Materia Condensada, Universidad Autónoma de Madrid, 28049 Madrid, Spain

[6] Present address: Max Planck Institute for Chemical Physics of Solids, 01187 Dresden, Germany

[*]e-mail: saul.velez@uam.es (S.V.); pietro.gambardella@mat.ethz.ch (P.G.)

**Table of contents:**

**Supplementary Note 1. Structural and topographic characterization of YIG/TmIG and YIG/TmIG/Pt**

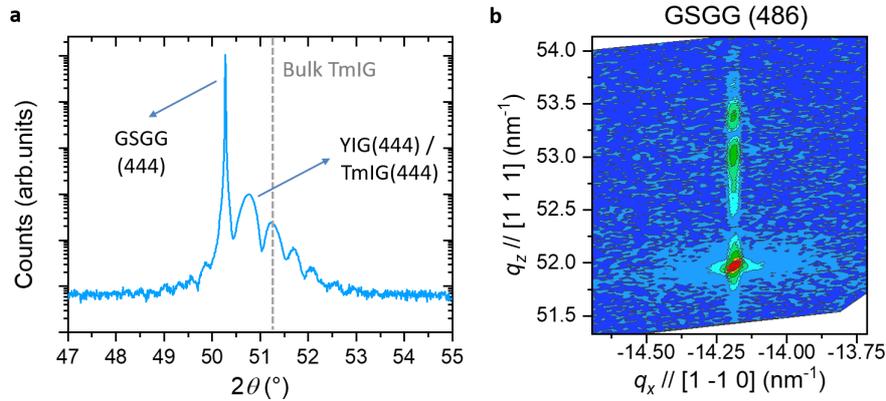

**Supplementary Figure 1 | Structural characterization of the YIG/TmIG films. a**, Symmetric X-ray diffraction scan of the GSGG/YIG(10 nm)/TmIG(20 nm) sample investigated in this work. As the X-ray response of a 10-nm-thick YIG is relatively weak[1], the signal in YIG/TmIG is dominated by the TmIG (444) diffraction peak and corresponding Laue oscillations. The peak shifts towards higher angles with respect to the bulk value[2] (dashed grey line) because of a reduction of the out-of-plane lattice constant due to tensile strain, in agreement with previous reports of TmIG films grown on GSGG[3,4]. **b**, Reciprocal space maps of the same sample around the GSGG (486) substrate peak ($q_x$ and $q_z$ are the in-plane and out-of-plane wavevectors along the crystal axes indicated). The in-plane lattice constants of the films and the substrate along the [1 -1 0] direction coincide, confirming full epitaxy. The colour code indicates the intensity of the diffraction peaks, with red (blue) corresponding to maximum (minimum) intensity.

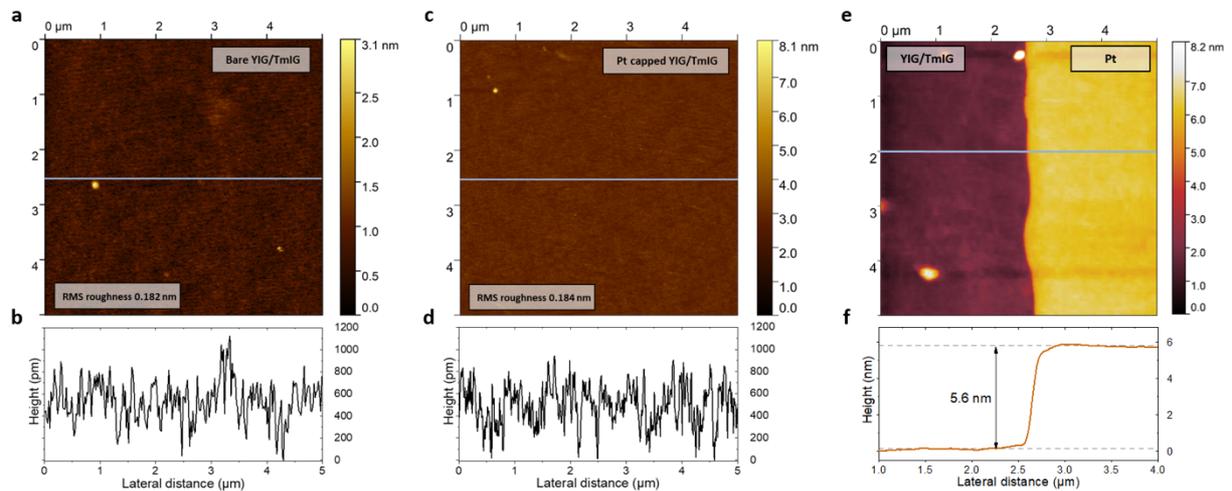

**Supplementary Figure 2 | Topographic characterization of YIG/TmIG and YIG/TmIG/Pt. a**, **c**, Atomic force microscopy (AFM) characterization of the surface topography of a bare YIG/TmIG and a Pt-capped YIG/TmIG heterostructure, respectively. The root mean square (RMS) roughness over a ~5 × 5 µm² area is on the order of 0.182 and 0.184 nm, respectively. **b**, **d**, AFM line profiles along the blue lines indicated in **a** and **c**, respectively, confirming the low roughness of our YIG/TmIG and YIG/TmIG/Pt films. **e**, AFM scan of the sample studied in the main text in a region partially covered with Pt. **f**, Average topography scan of the Pt edge along the blue line indicated in **e**, showing that the step size is about 5.6 nm. Consequently, we estimate the etching of the TmIG thickness in ~0.5 nm. The RMS surface roughness is below 0.2 nm on both the etched and Pt covered regions.

**Supplementary Note 2. Magnetic characterization of YIG/TmIG**

The magnetic anisotropy and saturation magnetization of the YIG(10nm)/TmIG(20nm)/Pt sample investigated in this work have been characterized by a combination of superconducting quantum interference device (SQUID) magnetometry and spin Hall magnetoresistance (SMR) measurements[3,5,6].

**Magnetic anisotropy**. Supplementary Figure 3 shows that the TmIG layer exhibits perpendicular magnetic anisotropy, with an effective anisotropy field $H_k \sim 1.4$ kOe, and that the YIG layer exhibits an easy-plane anisotropy with a part of the film rotating towards out-of-plane due to the exchange coupling with TmIG.

The smaller (larger) magnetic field required for saturating the magnetization in the out-of-plane (in-plane) configuration indicates that the dominant magnetic anisotropy of YIG/TmIG is out of plane (Supplementary Figs. 3a,3b). As SMR is only sensitive to the magnetic moments at the metal/insulator interface, the magnetic anisotropy of TmIG can be directly proven via transport measurements[3,4]. Supplementary Fig. 3c demonstrates that the magnetization of TmIG points out of plane at zero field, and gradually cants towards the plane as the in-plane field increases. From these measurements, we determine that the magnetic anisotropy of the bilayer is dominated by the perpendicular magnetic anisotropy of TmIG, which corresponds to an anisotropy field $H_k \sim 1.4$ kOe.

The small magnetic moment of the hysteresis loop for the in-plane measurement ($\sim 1.2 \times 10^{-8}$ A m$^2$ at zero field, Supplementary Fig. 3b) relative to the saturation magnetic moment of the full heterostructure ($\sim 7.7 \times 10^{-8}$ A m$^2$, Supplementary Fig. 3a) indicates that part of the YIG film magnetization lies in the plane of the film. This is not surprising, as YIG on GSGG is expected to exhibit in-plane anisotropy. From comparing the in-plane and out-of-plane data, we estimate that the magnetization of the first $\sim 3$ nm of YIG on GSGG lies in the plane of the film, while the rest gradually rotates towards out of the plane due to the exchange coupling with TmIG. The gradual increase of the magnetic moment with field above $H \sim 70$ Oe, Supplementary Fig. 3b, is consistent with the in-plane magnetic field gradually canting the magnetic moments of both the exchange-coupled YIG and the TmIG layer towards the plane, eventually achieving full saturation at $\sim 1.4$ kOe (Supplementary Fig. 3c; the relatively large paramagnetic response of GSGG prevents us to extract the saturation field from SQUID measurements). Importantly, the magnetic jump observed around zero field for the in-plane configuration indicates that the bottom part of the YIG film can be oriented with relatively small in-plane fields ($\sim 5$ Oe; see inset of Supplementary Fig. 3b). This allows for controlling the sign of the exchange field between YIG and TmIG, a result that is in agreement with the ratchet effect presented in Figs. 5 and 6 of the main text as well as Supplementary Notes 9 and 10.

**Saturation magnetization**. From the measurements shown in Supplementary Fig. 3, we estimate that the saturation magnetic moment of YIG(10nm)/TmIG(20nm) is $\sim 7.7 \times 10^{-8}$ A m$^2$. Taking into account that the surface area of the films is $\sim 26$ mm$^2$, and assuming that the saturation magnetization of the YIG film is $M_s(\text{YIG}) \sim 175$ kA m$^{-1}$ (Ref. [7]), we estimate the saturation magnetization of TmIG to be $M_s(\text{TmIG}) \sim 60$ kA m$^{-1}$.

**Temperature dependence**. The saturation magnetization reduces by about 18% when increasing the temperature from 300 to 350 K (Supplementary Fig. 4). We thus estimate that the TmIG magnetization decreases by a maximum of 7% due to Joule heating in the current-induced skyrmion dynamics experiments (see Extended Data Fig. 4 for the analysis of the current-induced Joule heating).

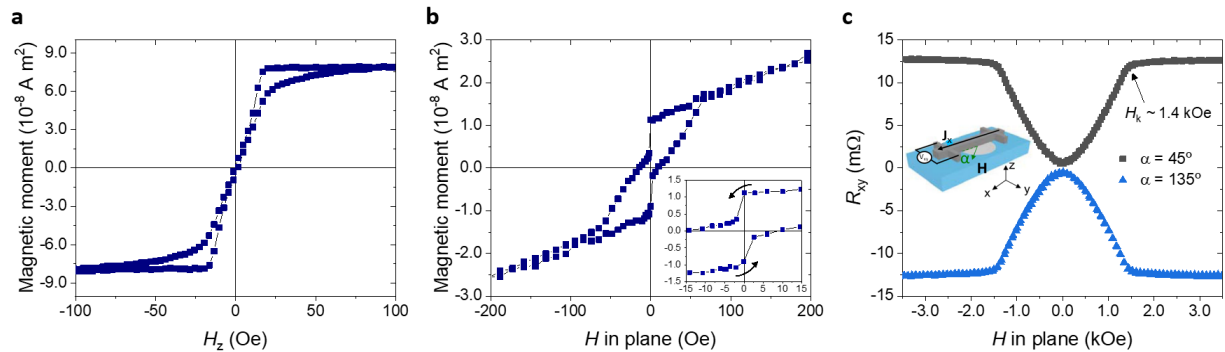

**Supplementary Figure 3 | Magnetic characterization of YIG(10nm)/TmIG(20nm). a,** Magnetic moment of the heterostructure as a function of out-of-plane field $H_z$. Both layers are fully saturated out of plane at $H_z \sim 100$ Oe. Note that the coercive field is smaller than the one shown in Fig. 1b of the manuscript. The larger coercive field in Fig. 1b is due to the pinning of domain walls at the device edges, resulting in a broadening of the hysteresis loop at the device area[3]. **b,** Magnetic moment as a function of in-plane field. *Inset*, magnification of the loop around zero field. The paramagnetic response of the GSGG substrate has been subtracted in **a** and **b**. **c,** Transverse SMR measurements as a function of in-plane field applied at an angle $\alpha$ with respect to the current direction. From these measurements, we can extract the magnetic anisotropy of TmIG[3,5,6]. At $H = 0$, $R_{xy} \approx 0$, indicating that the local magnetic moments of TmIG point out of the plane and are mostly randomly oriented, which is consistent with the data in **a** and the bottom images of Fig. 1c,d of the manuscript. As $H$ increases, the magnetization of TmIG cants towards the plane, resulting in a change of the amplitude of $R_{xy}$, which is maximum at $\alpha = 45°$ (positive change) or $135°$ (negative change). The saturation of $R_{xy}$ above $H_k \sim 1.4$ kOe indicates that the magnetic moments of TmIG are saturated in-plane, thus identifying $H_k$ as the anisotropy field of TmIG. We remark that $H_k$ is about half the value measured for single-layer TmIG films of the same thickness on GSGG[3,4], evidencing the role of the exchange coupling with YIG on the magnetic anisotropy of TmIG. Note that each measurement consists of superposed forward and backward field sweeps, indicating that the canting of the magnetic moments of TmIG does not exhibit hysteresis. A device-dependent constant offset is subtracted in **c** for convenience.

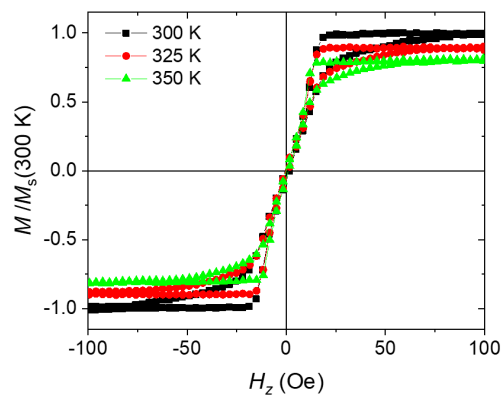

**Supplementary Figure 4. Temperature dependence of the out-of-plane magnetization**. The data are normalized to the saturation magnetization $M_s$ at 300 K.

**Supplementary Note 3. Chirality determination via nitrogen-vacancy magnetometry**

We used nitrogen-vacancy (NV) scanning magnetometry to characterize isolated skyrmion bubbles. This technique measures the stray field $B_{NV}$ produced by the magnetic textures of the sample at the position of the NV center with high field and spatial resolution, from which the spin texture of domain walls and skyrmions can be determined[3,8,9]. Figures 2a,d of the manuscript and Supplementary Fig. 6a show representative stray field scanning maps $B_{NV}(X, Y)$ of bubble domains located in YIG/TmIG/Pt and of a stripe domain intersecting a region partially covered by Pt.

**Skyrmion bubbles modelling.** We model the skyrmion as a closed ring of ellipsoidal shape, arbitrary orientation $\beta$, diametral axes $a$ and $b$, center $X_0, Y_0$, and wall width $\Delta_{DW}$. By adapting previous modelling of straight domain walls[3] to the bubble case, the domain wall profile is described by

$$M_X(r) = M_s \frac{\cos \psi}{\cosh \left( \dfrac{r - r_0}{\Delta_{DW}} \right)}$$

$$M_Y(r) = M_s \frac{\sin \psi}{\cosh \left( \dfrac{r - r_0}{\Delta_{DW}} \right)},$$

$$M_Z(r) = -M_s \tanh \left( \frac{r - r_0}{\Delta_{DW}} \right), \tag{1}$$

where $r_0$ indicate the position of the center of the wall for a given position along the wall ring of the bubble, $r$ the position in the direction perpendicular to the domain wall for the corresponding $r_0$, and $\psi$ is the chiral angle that describes the magnetic texture of the wall.

To fit the data, we normalized $B_{NV}(X, Y)$ to the maximum value to remove the influence of YIG on the stray field of TmIG. The fitting procedure is done by first finding the best domain wall width $\Delta_{DW}$ for a given domain wall type (Bloch, righ-handed Néel, and left-handed Néel), leaving $a$, $b$, and $\beta$ as free parameters. In a second step, $\Delta_{DW}$ is fixed and $a$, $b$, and $\beta$ (if $a \neq b$) are fitted. As an example, Fig. 2b of the manuscript shows the best fit of the stray field data presented in Fig. 2a, which corresponds to a circular bubble with right-handed Néel chirality, $\Delta_{DW} = 60$ nm, and $a \sim b = 950$ nm. The accuracy of the fitting is computed from the residual sum of squares (RSS) using following formula

$$\ln \mathcal{L}_1 - \ln \mathcal{L}_2 = -\frac{n}{2} \ln \frac{RSS_1}{RSS_2}, \tag{2}$$

where 1 and 2 indicate different sets of fit parameters with 2 being the one with smallest RSS, and $n$ the number of data points. Therefore, by using Supplementary Eq. (2) with different fit parameters, one can estimate the likelihood of a given type of domain wall. Supplementary Fig. 5 presents the likelihood of fitting the skyrmion data of Fig. 2a of the manuscript to different domain wall types characterized by $\psi$ and $\Delta_{DW}$. The likelihood plot clearly shows that the right-handed Néel chirality is the spin texture that best describes the skyrmions for wall widths in the range from $\Delta_{DW} \sim 30$ to $\sim 100$ nm, with best fit obtained with $\Delta_{DW} = 60$ nm. Taking into account the magnetic anisotropy of the film (Supplementary Note 2) and previous characterization of the domain wall width in TmIG[3], we expect the domain wall width to be about 50 nm. We thus conclude that in YIG/TmIG/Pt the domain wall texture is of right-handed Néel type. The same fitting procedure was followed for the deformed skyrmion presented in Fig. 2d-f of the manuscript.

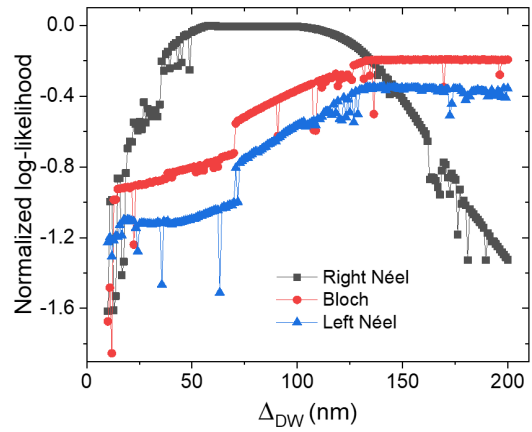

**Supplementary Figure 5 | Log-likelihood of the fits of the data of Fig. 2a** computed using Supplementary Eq. (2). Different $\Delta_{DW}$ values from 10 to 200 nm and different domain wall textures (Bloch, right-handed Néel, and left-handed Néel) were considered.

**Analysis of straight domain walls in YIG/TmIG/Pt and YIG/TmIG.** To infer the contribution of the YIG/TmIG interface to the DMI, we performed measurements of the stray field of a narrow stripe domain running across a region partially covered by Pt (Supplementary Fig. 6a). Direct inspection reveals that the stray field in the Pt-covered region is stronger than in the Pt-free region (Supplementary Fig. 6b). However, as we demonstrated in an earlier work, Pt also contributes to the stray field due to the magnetization induced by proximity with TmIG[3]. Therefore, for comparing the data, we subtracted the stray field associated to the Pt polarization in the YIG/TmIG/Pt region. The normalized line scans are very similar (Supplementary Fig. 6c), indicating that the domain wall type in YIG/TmIG is also right-handed Néel (as determined to be for YIG/TmIG/Pt; Fig. 2 and Supplementary Fig. 5) or of intermediate right-handed Néel-Bloch favored by a positive DMI induced by the YIG interface. Note that changes in $\Delta_{\mathrm{DW}}$ between the Pt-capped and Pt-free regions are expected to be negligible[10], and thus no significant influence of $\Delta_{\mathrm{DW}}$ on the stray field is expected. The relatively large uncertainty of the stray field data, however, does not allow us to conclude on the precise value of $\psi$. Nevertheless, the right-handed chirality in GSGG/YIG/TmIG is clearly different from that of TmIG directly grown on GSGG, which presents negative DMI and left-handed Néel domain walls[3]. As the TmIG/Pt interface has a positive DMI[3], we conclude that both YIG and Pt interfaces contribute to stabilize right-handed Néel domain walls and skyrmions in TmIG with an overall DMI strength above $D_{\mathrm{th}}$.

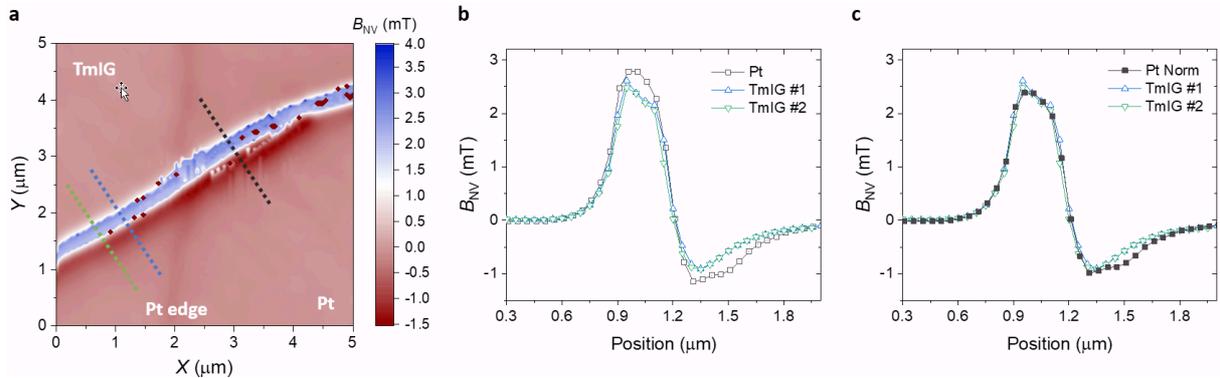

**Supplementary Figure 6 | Comparison of the stray field of a stripe domain in a region partially covered by Pt. a**, Stray field map $B_{\mathrm{NV}}(XY)$ of a stripe domain running across a region partially covered with Pt. Note that the stray field arises from two parallel domain walls. $|H_z| = 15$ Oe. **b**, Line scans of $B_{\mathrm{NV}}$ along the dashed lines in **a** (open symbols; the colour code identifies the line scan). **c**, Same as in **b** with the stray field in the Pt region corrected by the contribution of the Pt polarization to the stray field[3] (solid black symbols).

## Supplementary Note 4. Determination of the effective $g$ factor of TmIG

To estimate the effective g factor of TmIG we used the Wangsness relation[11,12]

$$\frac{M_{Fe}}{g_{Fe}} - \frac{M_{Tm}}{g_{Tm}} = \frac{M_s}{g},$$ (3)

where $M_{Fe}$ and $g_{Fe}$ are the magnetic moment and g factor of the $Fe^{3+}$ tetrahedral/octahedral sublattices, $M_{Tm}$ and $g_{Tm}$ the magnetic moment and g factor of the $Tm^{3+}$ dodecahedral sublattice, and $M_s = M_{Fe} - M_{Tm}$ and $g$ the net magnetic moment and effective g factor of TmIG. The negative sign accounts for the antiferromagnetic coupling between the $Fe^{3+}$ and $Tm^{3+}$ sublattices with both $M_{Fe}$ and $M_{Tm}$ defined positive and $M_{Fe} > M_{Tm}$. We take $g_{Fe} = 2$ (Ref. 7) and estimate $g_{Tm}$ from the expected $g$ factor of a free $Tm^{+3}$ ion. At the lowest spin-orbit multiplet state, the total angular momentum of $Tm^{3+}$ is $J = 6$ with an orbital momentum $L = 1$ and spin state $S = 5$, resulting in $g_{Tm} = 7/6$. The saturation magnetization of TmIG is estimated to be $M_s \sim 60$ kA m⁻¹ (see Supplementary Note 2).

In thin films, $M_{Fe}$ and $M_{Tm}$ may substantially deviate from the bulk values due to strain and finite size effects[1,3,13], giving a wide range of possible $M_{Fe}$, $M_{Tm}$ values for the solution of Supplementary Eq. (3). Supplementary Fig. 7 shows the value of $g$ computed by using Supplementary Eq. (3) and considering different combinations of $M_{Fe}$ and $M_{Tm}$ values. $g_{Fe}$ and $g_{Tm}$ are constrained to be 2 and 7/6, respectively. The dashed line indicates combinations with $M_{Fe} - M_{Tm} = 60$ kA m⁻¹. The bluish-coloured area corresponds to solutions with negative $g$ values, which is the case expected for our TmIG film according to the sign of the skyrmion Hall effect (Fig. 3 of the main text). By fixing $M_{Fe}$ to be 175 kA m⁻¹ (Ref. 7), we estimate $M_{Tm} \sim 115$ kA m⁻¹ and $g \sim -5.4$ (solution indicated by a blue dot in Supplementary Fig. 7).

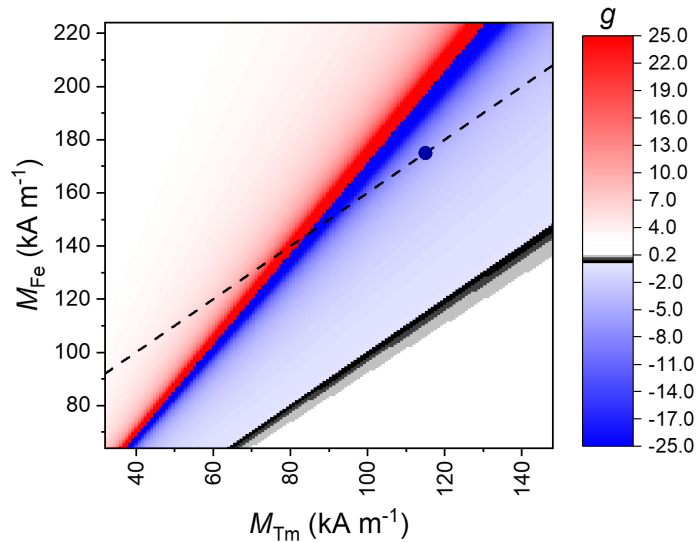

**Supplementary Figure 7 | Computed $g$ values of TmIG for different $M_{Fe}$ and $M_{Tm}$ combinations**. Computed $g$ values by using Supplementary Eq. (3) and $g_{Fe} = 2$ and $g_{Tm} = 7/6$. The dashed line indicates $M_{Fe}$, $M_{Tm}$ combinations with constant $M_s = 60$ kA m⁻¹. The bluish region corresponds to solutions with $g < 0$. The blue dot indicates the combination $M_{Fe} = 175$ kA m⁻¹ and $M_{Tm} = 115$ kA m⁻¹, which results in $g = -5.4$. See text for more details.

**Supplementary Note 5. Thiele's equation of a ferrimagnet and skyrmion velocity**

**Thiele's equation**. Under the approximation of point-like massless objects, the dynamics of skyrmions driven by current pulses is described by the modified Thiele's equation[14,15,16]

$$\mathbf{G} \times \mathbf{v}_{sk} - \alpha \boldsymbol{\mathcal{D}} \mathbf{v}_{sk} + \mathbf{F}_{SOT} = 0, \tag{4}$$

where $\mathbf{G} = G\hat{\mathbf{z}}$ is the gyromagnetic vector, $\mathbf{v}_{sk} = v_x \hat{\mathbf{x}} + v_y \hat{\mathbf{y}}$ the skyrmion velocity, $\alpha$ the damping parameter, $\boldsymbol{\mathcal{D}}$ the dissipative tensor, and $\mathbf{F}_{SOT} = F_{SOT}\hat{\mathbf{x}}$ the SOT driving force generated by $\mathbf{J}_x = J_x \hat{\mathbf{x}}$ (see Supplementary Fig. 8 for the schematics of the forces). $\hat{\mathbf{x}}$, $\hat{\mathbf{y}}$, and $\hat{\mathbf{z}}$ are unit vectors along the $x$, $y$, and $z$ directions, respectively. The gyromagnetic vector is given by

$$G = -4\pi \frac{M_s t}{\gamma} Q, \tag{5}$$

where $M_s$, $t$, and $\gamma = g \frac{\mu_B}{\hbar}$ are the saturation magnetization, thickness, and gyromagnetic factor of the magnetic layer, $g$ the Landé g factor, and $Q = \frac{1}{4\pi} \iint \left\{ \left( \frac{\partial \mathbf{m}}{\partial x} \times \frac{\partial \mathbf{m}}{\partial y} \right) \cdot \mathbf{m} \right\} dx dy$ the topological charge with $\mathbf{m}(x, y)$ the magnetic moment at position $(x, y)$. $\mu_B$ and $\hbar$ are the Bohr magneton and the reduced Planck constant.

**Gyromagnetic vector in a ferrimagnet.** In a ferrimagnet such as TmIG, the effective $\gamma$ can be computed from the saturation magnetization $M_{s,i}$ and the gyromagnetic factors $\gamma_i$ of the constituent sublattices (see Supplementary Eq. (3)), leading to[11,17]

$$\gamma = \frac{M_s}{\frac{M_{s,1}}{\gamma_1} - \frac{M_{s,2}}{\gamma_2}}. \tag{6}$$

In our TmIG films $\gamma < 0$ (see Supplementary Note 4), resulting in $G > 0$ for $Q = +1$ skyrmions. Note that Supplementary Eq. (5) can be rewritten as $G = 4\pi t s_{net} Q$ with $s_{net} = -\left( \frac{M_{s,1}}{\gamma_1} - \frac{M_{s,2}}{\gamma_2} \right) = -\frac{M_s}{\gamma}$.

**Skyrmion Hall angle**. The skyrmion deflection angle $\phi_{sk}$ induced by the Magnus force $\mathbf{G} \times \mathbf{v}_{sk}$ is given by[14,16,18]

$$\phi_{sk} \sim \tan^{-1} \left( \frac{G}{\alpha' \mathcal{D}} \right), \tag{7}$$

where the components of the dissipative tensor $\boldsymbol{\mathcal{D}}$ are given by[14,16] $\mathcal{D}_{ij} = -s_{net} t \iint \left\{ \frac{\partial \mathbf{m}}{\partial i} \cdot \frac{\partial \mathbf{m}}{\partial j} \right\} dx dy$, with $\mathcal{D}'_{12} = \mathcal{D}'_{21} = 0$ and $\mathcal{D}'_{11} = \mathcal{D}'_{22} \approx -s_{net} t \frac{2\pi R}{\Delta_{DW}}$ under the approximation $R \gg \Delta_{DW}$, with $R$ and $\Delta_{DW}$ being the radius of and domain wall with of the skyrmion bubble[19,20]. $\alpha'$ is the effective damping and is given by[17]

$$\alpha' = \alpha \frac{s_{tot}}{s_{net}}, \tag{8}$$

with $s_{tot} = -\left( \frac{M_{s,1}}{\gamma_1} + \frac{M_{s,2}}{\gamma_2} \right)$ the total angular momentum. By rewriting Supplementary Eq. (7) using the relations given above we obtain $\phi_{sk} \sim \tan^{-1} \left( -\frac{s_{net}}{s_{tot}} \frac{2Q\Delta}{\alpha R} \right)$ as given in Eq. (1) of the main text. Note that because $\frac{s_{\text{net}}}{s_{\text{tot}}}$ is negative in TmIG, $\phi_{\text{sk}}$ is positive (negative) for $Q = +1(-1)$ skyrmion bubbles as opposed to skyrmions in ferromagnets (see Supplementary Fig. 8 and Fig. 3 of the main text).

**Skyrmion velocity.** The skyrmion velocity $\boldsymbol{v}_{\text{sk}} = v_x \hat{\boldsymbol{x}} + v_y \hat{\boldsymbol{y}}$ in the flow regime is given by[16,21]

$$v_x = \frac{\eta}{1+\eta^2} \frac{F_{\text{SOT}}}{G}, \text{ and } v_y = \frac{1}{1+\eta^2} \frac{F_{\text{SOT}}}{G}, \tag{9}$$

and hence,

$$v_{\text{sk}} = |\boldsymbol{v}_{\text{sk}}| = \frac{1}{\sqrt{1+\eta^2}} \frac{F_{\text{SOT}}}{G}, \tag{10}$$

where

$$\eta = \frac{\alpha' D}{G} = -\frac{s_{\text{tot}}}{s_{\text{net}}} \frac{\alpha R}{2\Delta_{\text{DW}} Q} \text{ and } \frac{F_{\text{SOT}}}{G} = -\xi_{\text{DL}} J_x \gamma \frac{\pi R}{4}, \tag{11}$$

with $\xi_{\text{DL}}$ the effective field (per unit current density) associated to the damping-like SOT. Considering the values of $\xi_{\text{DL}}$ reported for TmIG/Pt (Refs. 4,22), we estimate a skyrmion mobility $\eta = v_{\text{sk}}/J_x$ exceeding $3 \times 10^{-8}$ m$^3$ A$^{-1}$ s$^{-1}$ in our TmIG devices, a value that is comparable to the mobility of ferrimagnetic domain walls near the angular momentum compensation[3,17,22,23,24].

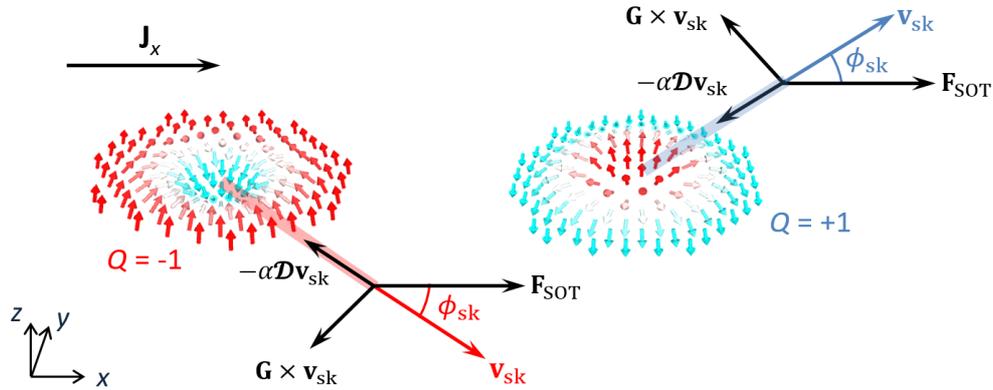

**Supplementary Figure 8 | Schematic of the skyrmion Hall effect in TmIG.** Schematics of $Q = -1$ and $Q = +1$ right-handed Néel skyrmion bubbles in TmIG and of the forces acting on the skyrmions due to the application of current pulses. The coordinate system and the direction of the current ($\mathbf{J}_x$), the skyrmion velocity ($\mathbf{v}_{\text{sk}}$), the sign of the deflection angle ($\phi_{\text{sk}}$), and the forces acting on the skyrmions (see Supplementary Eq. (4)) are indicated. The vectors indicate the direction of the magnetic moments. As the sign of $\mathbf{G}$ is positive (negative) for $Q = +1$ ($-1$) skyrmion bubbles, the sign of $\phi_{\text{sk}}$ in TmIG is opposite to the one encountered in ferromagnetic materials.

**Supplementary Note 6. Magnetic field dependence of the skyrmion radius.**

The average skyrmion radius is analyzed from MOKE measurements taken on isolated bubbles and for different values of $H_z$ (Extended Data Fig. 1). We note that the average radius inferred from the MOKE analysis (Extended Data Fig. 1c) is consistent with the radius extracted from NV magnetometry measurements (Fig. 2a-c of the manuscript). Both the increase of the bubble radius (Extended Data Fig. 1c) and the destabilization of the bubble domains into stripe domains when decreasing $|H_z|$ (Fig. 1d of the manuscript) are a consequence of the minimization of the magnetostatic energy of the bubble with field[25]. For fields below $|H_z| \lesssim 10$ Oe we cannot find only skyrmion bubbles by either current or field sweep protocols as the bubbles tend to expand and transform into stripe domains.

**Supplementary Note 7. Skyrmion deformations due to pinning**

**Skyrmion ellipticity.** The statistical analysis of the bubbles' shape extracted from MOKE data show that most skyrmion bubbles present an ellipticity $b/a \gtrsim 0.9$ (Extended Data Figs. 1a and 2a), indicating that the bubbles tend to retain a circular shape after a current pulse. $a$ and $b$ define the larger and smaller axes of the ellipsoid. Such ellipsoidal deformations correspond to relative contractions/elongations of the skyrmion diameter of about $50 \pm 50$ nm between each other. These results are consistent with an average distance between pinning centers of about 50 to 100 nm if the deformations are assumed to arise from the hopping of the skyrmion wall between two adjacent defects.

**Current-induced bubble deformations.** We found a correlation between the most preferred direction of bubble deformation relative to the direction of the current pulses. Concretely, we found that the bubbles exhibit a larger probability to exhibit deformation in the direction of motion as well as perpendicular to it (Extended Data Fig. 2b). Whereas the former indicates that the deformations correlate with the direction of skyrmion motion set by the skyrmion Hall effect and pinning[20,26,27], the latter is consistent with bubble distortions induced by SOTs[28].

**Supplementary Note 8. Alternative explanations for the pulse length dependence: inertial and automotion effects**

**Inertial effects.** A possible explanation for the finite bubble displacements observed as $t_{\mathrm{p}} \to 0$ (Fig. 4c,d of the main text as well as Extended Data Figs. 5 and 6) is that the skyrmion bubbles behave as objects with a finite mass and thus keep moving after the end of a current pulse. This effect would be detected in MOKE experiments if there is an asymmetry in the acceleration and deceleration times of the skyrmion bubbles, as reported for Néel domain walls in low-damping media and moderate DMI[29]. However, such inertia effects are expected to emerge for domain walls and skyrmions driven in the flow regime. In the presence of random hopping produced by disorder and thermal fluctuations, as observed in our experiments for skyrmion bubbles driven in the creep regime (Figs. 4 and 5 of the main text), we expect that the inertia effects would be negligible.

**Automotion.** Another explanation for the pulse length dependence of $\overline{\Delta x}$, $\overline{\Delta y}$ and $\bar{v}_{\mathrm{sk}}$ (Fig. 4c-e of the main text) is that the skyrmion bubbles exhibit automation effects as the result of the displacement of vertical Bloch lines around the bubble boundary, in analogy with the behaviour of magnetic bubbles in a field gradient observed in thick garnet layers[16,30,31]. The displacement of vertical Bloch lines would be driven by the reversal of $\mathbf{M}_{\mathrm{YIG}}$ induced by the in-plane Oersted fields generated by the pulses ($\mathbf{H}_{\mathrm{Oe},y} \propto +J_{\mathrm{x}}\hat{y}$ at the YIG plane), and therefore inertia effects may only emerge for one polarity of $\mathbf{J}_{\mathrm{x}}$ for a given $\mathbf{M}_{\mathrm{YIG}}||\mathbf{y}$ configuration. This scenario, however, is ruled out because a similar behaviour is observed for all orientations of $\mathbf{M}_{\mathrm{YIG}}$ and $\mathbf{J}_{\mathrm{x}}$. Extended Data Figure 6 shows representative data taken for $\mathbf{M}_{\mathrm{YIG}} = -M_{\mathrm{YIG}}\hat{y}$, revealing finite displacements for both polarities of $\mathbf{J}_{\mathrm{x}}$ as $t_{\mathrm{p}} \to 0$. In addition, analysis of the current threshold for bubble depinning as function of $\mathbf{J}_{\mathrm{x}}$ amplitude, in-plane field $H_{\mathrm{y}}$, and pulse length $t_{\mathrm{p}}$ suggests that the in-plane Oersted fields are not capable to produce significant changes to $\mathbf{M}_{\mathrm{YIG}}$ (Supplementary Note 9), which we ascribe to the relatively large thickness of TmIG and the moderate current densities employed in the experiments.

**Supplementary Note 9: Influence of the Oersted field and $H_y$ on the skyrmion dynamics**

As demonstrated in Figs. 5 and 6 of the manuscript, both the skyrmion depinning probability and the velocity of the skyrmion bubbles strongly depend on the orientation of $\mathbf{M}_{\mathrm{YIG}}$ relative to $\mathbf{J}_x$. To control the orientation of $\mathbf{M}_{\mathrm{YIG}}$, a small in-plane magnetic field $\mathbf{H}_y = H_y\hat{y}$ is applied (see Supplementary Note 2 for more details regarding the magnetic properties of the films; in particular, Supplementary Fig. 3b shows that an in-plane field as small as 3 Oe can significantly modify $\mathbf{M}_{\mathrm{YIG}}$). It is therefore crucial to determine whether the asymmetry in the dynamics of the skyrmion bubbles with $\mathbf{J}_x$ may arise from the influence of the in-plane Oersted field $\mathbf{H}_{\mathrm{Oe},y} = H_{\mathrm{Oe},y}\hat{y} \propto +J_x\hat{y}$ generated by the pulses on $\mathbf{M}_{\mathrm{YIG}}$, as well as to determine whether $\mathbf{H}_y$ itself influence the dynamics. To investigate these questions, we determined the current threshold $J_x^{\mathrm{th}}$ for skyrmion depinning as function of $\mathbf{J}_x$ orientation, pulse length $t_{\mathrm{p}}$, and $H_y$ strength.

Supplementary Figure 9 shows representative data taken for $H_y < 0$ and $Q = +1$ ($H_z = -20$ Oe). At magnetic fields below $|H_y|\sim 4$ Oe, the current threshold is rather independent on the magnetic field and the polarity of the current, but above that field value, a strong asymmetry with $\mathbf{J}_x$ emerges, with the current threshold becoming larger for $J_x < 0$, an asymmetry that is in agreement with the results presented in Fig. 6e,f of the main text. Moreover, when reversing the direction of $H_y$ (not shown here), we observe that the current threshold becomes larger for $J_x > 0$, while it stays rather constant for $J_x < 0$, also in agreement with the asymmetry with $\mathbf{J}_x$ for $\mathbf{M}_{\mathrm{YIG}} = +M_{\mathrm{YIG}}\hat{y}$ presented in Fig. 6g,h of the main text (see Supplementary Note 10 for more details regarding the asymmetries of the ratchet effect). We note that the results shown in Supplementary Fig. 9 are not dependent on the particular sequence followed with the magnetic field $\mathbf{H}_y$ before starting the measurements, suggesting that the application of current pulses randomize the domains towards the equilibrium configuration set by the external field for values in the range $|H_y| < -10$ Oe.

The rather constant $J_x^{\mathrm{th}}$ with $H_y < 0$ observed for $J_x > 0$ (Supplementary Fig. 9b) is attributed to the fact that no significant difference should be observed in $J_x^{\mathrm{th}}$ between a demagnetized case ($H_y = 0$ Oe) and a saturated one along the favoured $\mathbf{M}_{\mathrm{YIG}}$ direction. That is because a demagnetized case presents domains with both favoured and unfavoured $\mathbf{M}_{\mathrm{YIG}}$ orientations as well as intermediate ones aligned with the current. The average skyrmion velocity for $J_x > 0$, however, increases when $\mathbf{M}_{\mathrm{YIG}}$ saturates along $-y$, in agreement with the data shown in Figs. 4a,b and 5a,b of the main text. Further, the velocity is found to be weakly dependent on the external field for fields from $|H_y| \sim -10$ Oe to $\sim -25$ Oe, the later defining the threshold for the destabilization of the bubble domains into stripe domains.

We now focus our attention on the field dependence of the current threshold for $J_x < 0$ and $H_y < 0$ (Supplementary Fig. 9a). Remarkably, while the depinning current density increases by a factor $\sim 4$ when decreasing the pulse length from 100 to 15 ns (blue up triangles and green down triangles, respectively; note that the $t_{\mathrm{p}}$-dependence of $J_x^{\mathrm{th}}$ is in agreement with the data presented in Extended Data Fig. 3), the field-dependence remains qualitatively the same, with a field-independent regime observed from $|H_y| \sim 10$ to 20 Oe. The fact that the field independent regime is reached at the same value for all current conditions (despite the associated $H_{\text{Oe},y}$ field differs by a factor $\sim 4$ between the extreme cases: $J_x \sim 8.5 \times 10^{11}$ A m$^{-2}$ for $t_p = 15$ ns, while $J_x \sim 2.2 \times 10^{11}$ A m$^{-2}$ for $t_p = 100$ ns), indicates that the influence of $\mathbf{H}_{\text{Oe},y}$ on $\mathbf{M}_{\text{YIG}}$ is negligible and that the magnetization of the YIG film is saturated above $|H_y| = H_s = 10$ Oe (indicated by a vertical line), further indicating that $H_y$ has a negligible effect on the skyrmion dynamics compared to $\mathbf{M}_{\text{YIG}}$. We thus conclude that the ratchet effect arises from the exchange coupling of $\mathbf{M}_{\text{YIG}}$ with the skyrmions in TmIG.

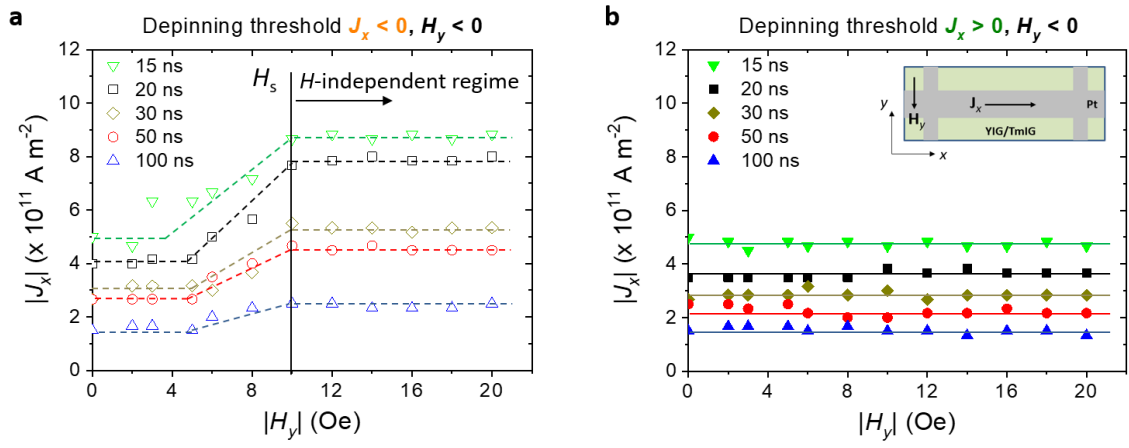

**Supplementary Figure 9 | Current threshold for the depinning of skyrmion bubbles as function of $\mathbf{J_x}$, $\mathbf{t_p}$, and $\mathbf{H_y}$. a**, Current threshold as function of $H_y$ for different pulse lengths. The polarity of the magnetic field and the current applied are $H_y < 0$ and $J_x < 0$, which corresponds to the unfavoured configuration for bubble motion. Same field dependence is observed for $H_y > 0$ and $J_x > 0$. The dashed lines are guides to the eye. A field-independent regime is reached at $H_s \sim 10$ Oe for all $t_p$, $|J_x|$ conditions (indicated with a vertical solid line). **b**, Same as **a**, but for $J_x > 0$ and $H_y < 0$, i.e., a favoured configuration for skyrmion motion. Same data is obtained for $J_x < 0$ and $H_y > 0$. The solid lines are guides to the eye. Data in **a** and **b** correspond to $Q = +1$ skyrmion bubbles with $H_z = -20$ Oe. Same behaviour for the current depinning threshold is observed for $Q = -1$ skyrmion bubbles and $H_z = +20$ Oe. The inset indicates the orientation of $\mathbf{J_x}$ and $\mathbf{H_y}$ relative to the current line. See Supplementary Note 10 and Figs. 5 and 6 of the main text for more details regarding the symmetries of the skyrmion dynamics with $\mathbf{M}_{\text{YIG}}$, $\mathbf{J_x}$, and $Q$.

**Supplementary Note 10. Skyrmion ratchet effect: supplementary data**

| $Q$ | $\mathbf{M_{YIG}}$ ($\hat{\boldsymbol{y}}$) | $\mathbf{J}_x$ ($\hat{\boldsymbol{x}}$) | $\mathbf{T^{DL}}$ ($\hat{\boldsymbol{y}}$) | $|\mathbf{v_{sk}}|$ | Ratchet ($\hat{\boldsymbol{x}}$) |
|---|---|---|---|---|---|
| +/− | Demag | +/− | | Intermediate | No |
| +/− | − | + | + | Fast | + |
| +/− | − | − | − | Slow or pinned | + |
| +/− | + | + | + | Slow or pinned | − |
| +/− | + | − | − | Fast | − |

**Supplementary Table 1 | Skyrmion dynamics with $\mathbf{M_{YIG}}$, $\mathbf{J}_x$, and $Q$: symmetry of the ratchet effect.** Comparison of the skyrmion dynamics for different $\mathbf{M_{YIG}}$, $\mathbf{J}_x$, and $Q$ configurations relative to the YIG demagnetized case. The vectors $\hat{\boldsymbol{x}}$ and $\hat{\boldsymbol{y}}$ indicate the orientation of the vectors $\mathbf{M_{YIG}}$, $\mathbf{J}_x$, $\mathbf{T^{DL}}$, and the ratchet effect, and the sign + or − their polarity (see Fig. 1a of the main text for the definition of the sample coordinates). Note that the polarity of $\mathbf{T^{DL}}$ is given by $\mathbf{J}_x$ and that here we only consider the sign of the $\hat{\boldsymbol{y}}$ component at the bubbles wall as is the relevant one for the ratchet effect. Further, we only consider configuration with $\mathbf{M_{YIG}}$ orthogonal to $\mathbf{J}_x$ as no asymmetry in the dynamics is observed when $\mathbf{M_{YIG}}$ and $\mathbf{J}_x$ are collinear (Extended Data Fig. 7). When the polarity of $\mathbf{J}_x(\hat{\boldsymbol{x}})$ and $\mathbf{M_{YIG}}(\hat{\boldsymbol{y}})$ are the same, the dynamics of the skyrmions are slow or pinned (combinations indicated by orange colour), but when they are opposite, the skyrmion motion is efficient and faster (indicated by green colour) relative to the demagnetized case. See also Figs. 5 and 6 of the main text, which present representative data of the dynamics of the skyrmion bubbles with $\mathbf{J}_x(\hat{\boldsymbol{x}})$ and $\mathbf{M_{YIG}}(\hat{\boldsymbol{y}})$. For a given $\mathbf{M_{YIG}}(\hat{\boldsymbol{y}})$ orientation, the asymmetry in the skyrmion dynamics with $\mathbf{J}_x(\hat{\boldsymbol{x}})$ leads to the ratchet effect indicated in the last column, with the bubble motion being preferred towards $+\hat{\boldsymbol{x}}$ for $\mathbf{M_{YIG}}$ aligned to $-\hat{\boldsymbol{y}}$, while motion is preferred towards $-\hat{\boldsymbol{x}}$ for $\mathbf{M_{YIG}}$ saturated along $+\hat{\boldsymbol{y}}$ (note that here the deflection of the skyrmion bubbles towards $\pm\hat{\boldsymbol{y}}$ due to the topological Hall effect is not considered for simplicity). The same asymmetric motion is observed for both $Q = +1$ and $-1$ skyrmion bubbles. As schematized in Fig. 6a-d of the main text, the asymmetric dynamics with $\mathbf{J}_x$ originates from the distortion of the magnetic configuration of the skyrmion bubbles produced by $\mathbf{T^{DL}}$ (which depends on $\mathbf{J}_x$) relative to the one induced by the exchange coupling with $\mathbf{M_{YIG}}$. When the distortions oppose (favour) each other, the net distortion towards $\hat{\boldsymbol{y}}$ becomes smaller (larger), resulting in stronger (weaker) $\mathbf{F_{SOT}}$ driving forces. See also Supplementary Fig. 10, which provide additional sketches of the expected distortion of the skyrmion bubbles for other $\mathbf{M_{YIG}}$, $\mathbf{J}_x$, and $Q$ configurations than the ones presented in Fig. 6a-d of the main text.

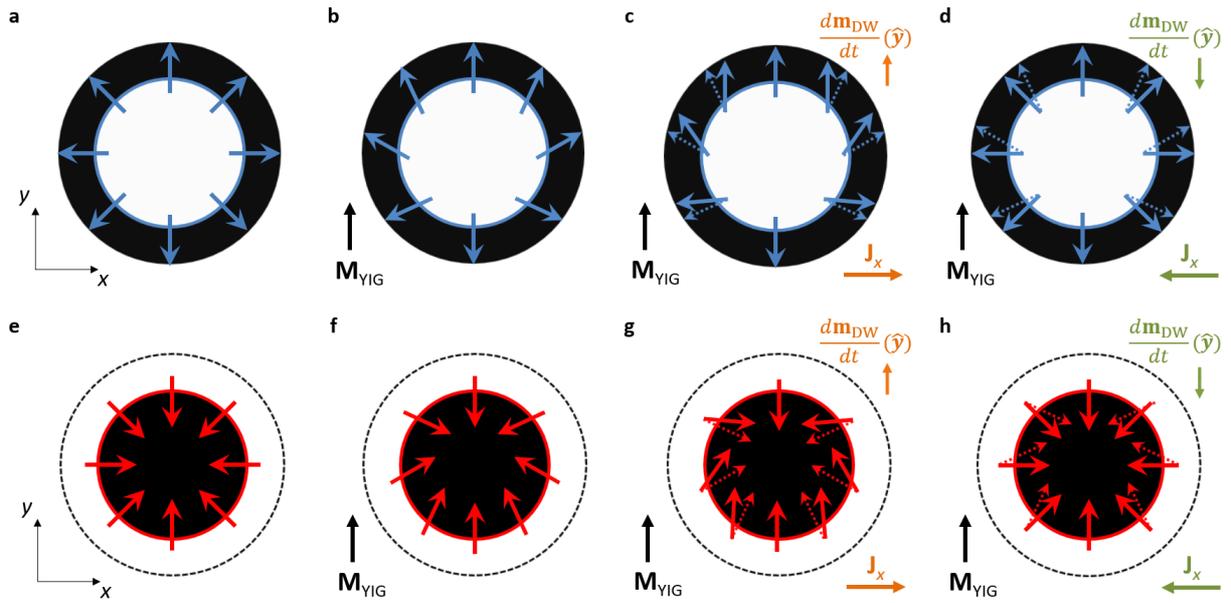

**Supplementary Figure 10 | Schematics of the magnetic distortion of the skyrmion bubbles with $\mathbf{M}_{\text{YIG}}$ and $\mathbf{J}_x$: additional $\mathbf{M}_{\text{YIG}}$ and $Q$ configurations. a-d,** Same as Fig. 6a-d of the main text, but for $\mathbf{M}_{\text{YIG}} = +M_{\text{YIG}}\hat{\mathbf{y}}$. As opposed to the case of Fig. 6a-d, the distortion of the skyrmion bubble is enhanced (reduced) for $J_x > 0$ ($J_x < 0$). Consequently, for $\mathbf{M}_{\text{YIG}} = +M_{\text{YIG}}\hat{\mathbf{y}}$ the skyrmion motion is more efficient for $J_x < 0$ than for $J_x > 0$, in agreement with the results presented in Fig. 6g,h of the main text. **e-h,** Same as **a-d,** but for a $Q = -1$ skyrmion bubble. The symmetry of the resulting torques with $\mathbf{J}_x$ is the same as for $Q = +1$ skyrmion bubbles, i.e., larger (smaller) distortions are observed for $J_x > 0$ ($J_x < 0$). See **c,d** and **g,h.** In the later, only the direction of the $\hat{\mathbf{y}}$ component of the $d\mathbf{m}_{\text{DW}}/dt$ induced by the torques is depicted for simplicity. Fast (slow) motion is expected when $d\mathbf{m}_{\text{DW}}/dt(\hat{\mathbf{y}})$ opposes $\mathbf{M}_{\text{YIG}}$.

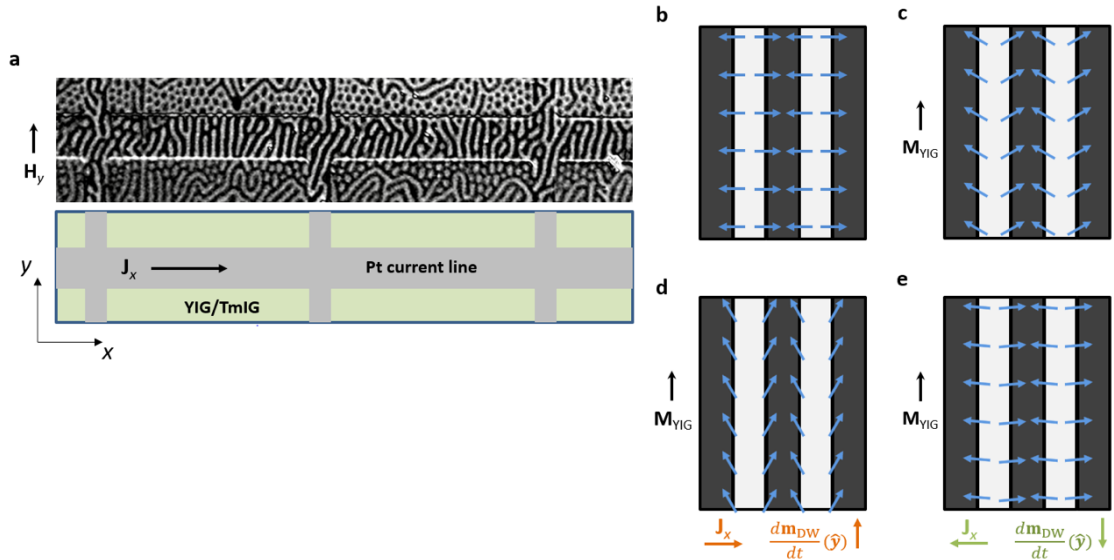

**Supplementary Figure 11 | Ratchet effect for stripe domains. a**, *Top*, Differential MOKE image showing stripe domains in TmIG. The white (dark) contrast indicates domains with the magnetization pointing up (down). An in-plane magnetic field $H_y = +30$ Oe is applied along $+\hat{y}$ to stabilize stripe domains oriented perpendicular to the current line. $H_z = -10$ Oe. *Bottom*, Schematics indicating the position of the current line in the MOKE image. When applying current pulses $\mathbf{J}_x$, as for the case of skyrmion bubbles with $\mathbf{M}_{YIG} = +M_{YIG}\hat{y}$, we observe that the dynamics of the stripe domains is more efficient for $J_x < 0$ than for $J_x > 0$ (see Fig. 6g,h of the main text). When the orientation of $\mathbf{H}_y$ is reversed, the asymmetry in the dynamics of the stripe domains with $\mathbf{J}_x$ is also reversed, agreeing with the asymmetry with $\mathbf{M}_{YIG}$ (which is set by $\mathbf{H}_y$) and $\mathbf{J}_x$ as summarized in Supplementary Table 1. The explanation of the effect is the same as for the case of skyrmion bubbles. The motion of the stripe domains is more (less) efficient when the distortion of the magnetic moments at the stripe's walls by $\mathbf{M}_{YIG}$ and $\mathbf{J}_x$ is minimal (maximal). See sketches from **b** to **e**. The bubble domains surrounding the current line are induced by the out of plane component of the Oersted field for currents $\mathbf{J}_x \gtrsim 8 \times 10^{11}$ A m$^{-2}$ at this field values. **b**, Schematics of the expected orientation of the magnetic moments at the walls of the stripe domains in TmIG (as discussed in the main text, the domain walls should exhibit right-handed Néel chirality). **c**, Magnetic distortion of the walls due to $\mathbf{M}_{YIG} = +M_{YIG}\hat{y}$. **d**, **e**, Additional magnetic distortion due to $\mathbf{J}_x$, showing that a more efficient wall motion is expected for $J_x > 0$ (**d**) than for $J_x < 0$ (**e**).